\documentclass[prd,onecolumn,superscriptaddress,floatfix,nopacs,preprintnumbers,nofootinbib]{revtex4}
\usepackage[utf8]{inputenc}

\usepackage{graphicx}
\usepackage[normalem]{ulem}
\usepackage{xcolor}

\usepackage{mathrsfs}
\usepackage{amssymb,bm}
\usepackage{amsmath}
\usepackage{mathtools}
\usepackage{physics}
\usepackage{slashed}
\usepackage{verbatim}

\newcommand{\nc}{N_\mathrm{c}}

\newcommand{\ft}{F_{2}}
\newcommand{\fl}{F_{\mathrm{L}}}
\newcommand{\nf}{n_\mathrm{f}}
\newcommand{\tf}{T_\mathrm{f}}
\newcommand{\seq}{\sum_{q}^{\nf} e_q^2}

\newcommand{\cf}{C_\mathrm{F}}
\newcommand{\TR}{T_\mathrm{R}}
\newcommand{\msbar}{\overline{\rm MS}}
\newcommand{\seqav}{ \bar{e}_q^2}

\newcommand{\cflglo}{ {C}_{\fl g}^{(1)}}
\newcommand{\cflqlo }{ {C}_{\fl q}^{(1)}}

\newcommand{\cflgnlo}{ {C}_{\fl g}^{(2)}}
\newcommand{\cflnsnlo }{ {C}_{\fl \rm NS}^{(2)}}
\newcommand{\cflpsnlo }{ {C}_{\fl \rm PS}^{(2)}}
\newcommand{\cftgnlo}{ {C}_{\ft g}^{(1)}}
\newcommand{\cftqnlo }{ {C}_{\ft q}^{(1)}}

\newcommand{\cfkqnlo }{ {C}_{\fk q}^{(1)}}
\newcommand{\wft}{\widetilde{F}_2}

\newcommand{\wftp}{\widetilde{F'}_2}

\newcommand{\wfl}{\widetilde{F}_{\mathrm{L}}}
\newcommand{\wflp}{\widetilde{F'}_{\mathrm{L}}}
\newcommand{\wflpp}{\widetilde{F''}_{\mathrm{L}}}
\newcommand{\ftwminus }{F_2^{\rm W^-}}

\newcommand{\ftwplus }{F_2^{\rm W^+}}

\newcommand{\ftwdelta }{\Delta F_2^{\rm W}}
\newcommand{\wftwdelta}{\Delta \widetilde{F}_2^{\rm W}}
\newcommand{\fk }{F_3}
\newcommand{\fkw }{F_3^{\rm W^-}}
\newcommand{\ftcw }{F_{2\rm c}^{\rm W^-}}
\newcommand{\wftcw}{\widetilde{F}_{2\rm c}^{\rm W^-}}
\newcommand{\xq }{ (x, Q^2)}
\newcommand{\eu }{ e_u^2}
\newcommand{\ed }{ e_d^2}
\newcommand{\es }{ e_s^2}

\newcommand{\PSlo }{ \Sigma^{(0)}}
\newcommand{\singletlo }{ \Xi^{(0)}}
\newcommand{\Glo }{ G^{(0)}}
\newcommand{\PSnlo }{ \Sigma^{(1)}}
\newcommand{\singletnlo }{ \Xi^{(1)}}
\newcommand{\Gnlo }{ G^{(1)}}
\newcommand{\Pqqlo }{ P_{qq}^{(0)}}
\newcommand{\Pqglo }{ P_{qg}^{(0)}}
\newcommand{\Pgqlo }{ P_{gq}^{(0)}}
\newcommand{\Pgglo }{ P_{gg}^{(0)}}
\newcommand{\Pqqnlo }{ P_{qq}^{(1)}}
\newcommand{\Pqgnlo }{ P_{qg}^{(1)}}
\newcommand{\Pgqnlo }{ P_{gq}^{(1)}}
\newcommand{\Pggnlo }{ P_{gg}^{(1)}}
\newcommand{\Pplusnlo }{ P_{qq}^{+(1)}}
\newcommand{\Pminusnlo }{ P_{qq}^{-(1)}}
\newcommand{\Psingletnlo }{ P_{qq}^{S(1)}}
\newcommand{\PVqqnlo }{ P_{qq}^{V(1)}}
\newcommand{\PVqqbarnlo }{ P_{q\overline{q}}^{V(1)}}

\usepackage[breaklinks,colorlinks,citecolor=citcolor,urlcolor=blue,linkcolor=lcolor]{hyperref}

\definecolor{lcolor}{rgb}{0.5,0,0}
\definecolor{citcolor}{rgb}{0,0.3,0.0}

\newcommand{\as}{\alpha_\mathrm{s}}
\newcommand{\As}{\frac{\alpha_\mathrm{s}}{2\pi}}

\begin{document}

\title{Next-to-leading order evolution of structure functions without PDFs}

\author{Tuomas Lappi}
\email{tuomas.v.v.lappi@jyu.fi}
\author{Heikki Mäntysaari}
\email{heikki.mantysaari@jyu.fi}
\author{Hannu Paukkunen}
\email{hannu.paukkunen@jyu.fi}
\author{Mirja Tevio}
\email{mirja.h.tevio@jyu.fi}
\affiliation{
University of Jyväskylä, Department of Physics,  P.O. Box 35, FI-40014 University of Jyväskylä, Finland
}
  
\affiliation{
Helsinki Institute of Physics, P.O. Box 64, FI-00014 University of Helsinki, Finland
}

\begin{abstract}
We formulate and numerically solve the Dokshitzer-Gribov-Lipatov-Altarelli-Parisi~(DGLAP) evolution equations at next-to-leading order in perturbation theory directly for a basis of 6 physical, observable structure functions in deeply inelastic scattering. By expressing the  evolution 
in the physical basis one evades the factorization scale and scheme dependence. Working in terms of observable quantities, rather than  parametrizing and fitting unobservable parton distribution functions (PDFs), provides an unambiguous way to confront predictions of perturbative Quantum Chromodynamics with experimental measurements. We compare numerical results for the DGLAP evolution for structure functions in the physical basis  to the conventional evolution with PDFs.

\end{abstract}

\maketitle

\section{Introduction}
\label{sec:intro}

Future experiments on deeply inelastic scattering (DIS) at the Electron-Ion Collider \cite{AbdulKhalek:2021gbh}, and later at the Large Hadron-electron Collider \cite{LHeC:2020van} and Future Circular Collider \cite{FCC:2018vvp} are anticipated to provide new DIS data within wider kinematical regions than before, supplementing the wide range of already available DIS measurements, like the ones in Refs.~\cite{H1:2015ubc,H1:2013ktq,ZEUS:2014thn,Berge:1989hr,CHORUS:2005cpn, NuTeV:2005wsg,NuTeV:2001dfo,NOMAD:2013hbk}. Especially interesting for understanding fundamental properties of Quantum Chromodynamics (QCD) with new high energy data is  the small-$x$ region, where in the regime of gluon saturation the linear Dokshitzer-Gribov-Lipatov-Altarelli-Parisi~(DGLAP) evolution \cite{Dokshitzer:1977sg,Gribov:1972ri,Gribov:1972rt,Altarelli:1977zs} approach is expected to reach its limit of validity.  
Interpreting the new experimental measurements requires, in addition to perturbative calculations at higher orders,  a renewed attention to systematic uncertainties in the theoretical approaches \cite{Cacciari:2011ze,David:2013gaa,Bagnaschi:2014wea,NNPDF:2019vjt}.

We will argue in this paper that in order to address questions on the validity of the collinear, perturbative-QCD (pQCD) picture, 
it is useful to parametrize the needed nonpertubative inputs, and formulate theoretical predictions, purely in terms of physical observables~\cite{Furmanski:1981cw}. Calculating physical observables, whether measured or not, is also the only unambiguous way to compare   different theoretical approaches~\cite{Armesto:2022mxy}.

In the usual collinear factorization framework \cite{Collins:1989gx} cross sections are factorized into a perturbatively calculated short distance part, and a long distance part described by parton distribution functions (PDFs) \cite{Gao:2017yyd,Kovarik:2019xvh,Ethier:2020way,Klasen:2023uqj}. The factorization scale $\mu_f$, which defines the separation between  the short and long distance physics, remains arbitrary. Typically, to avoid large logarithms of scale ratios, the factorization scale is taken to be given by a physical scale, such as the invariant mass off the exchanged boson in DIS, $\mu_f^2=Q^2$, times a constant of order one. However, any such choice still leaves some degree of arbitrariness in the calculation, requiring some prescription for how to estimate the associated theoretical uncertainties.  
Moreover, the result of a perturbative calculation at a fixed perturbative order depends on the factorization scheme, which is usually chosen to be the $\msbar$ scheme. The PDFs, as non-observable quantities, have a significant  parametrization freedom, which results in additional uncertainties. For decades various PDF collaborations have been establishing PDF sets, with some reaching as high as next-to-next-to-next-to-leading order \cite{McGowan:2022nag, NNPDF:2024nan} in the strong coupling $\as$. 
As the perturbative accuracy in terms of powers of $\as$ of calculations in the collinear factorization approach  has improved over the recent years, it becomes increasingly important to devote attention to systematic ways to address the remaining scheme, scale and parametrization uncertainties discussed above. In a recent publication~\cite{Delorme:2025teo} a systematic study of the uncertainty related to the choice of the factorization scheme in LHC phenomenology is carried out for the first time.

One potential approach to decrease the theoretical uncertainty is to replace the PDFs with physical quantities. In this alternative method, which we call the physical-basis approach, the DGLAP evolution is formulated directly in terms of the DIS structure functions. To achieve this, let us start 
from the definition of a general structure function in the collinear factorization framework:
\begin{equation}
\label{eq: structure function general notation}
    F_{i}\xq = \sum _j C_{ij}\left(\as(\mu_r^2), Q^2,\mu_f^2,\mu_r^2 \right)\otimes f_j(\mu_f^2)\,,
\end{equation}
were the $Q^2$ dependence appears in terms of logarithms $\log(Q^2/\mu_r^2)$ and $\log(Q^2/\mu_f^2)$, where $\mu_r^2$ denotes the renormalization scale.  The conventional DGLAP evolution for the PDFs $f_j(\mu_f^2)$ would be derived, using the independence of the physical observable  $F_{i}\xq$ of $\mu_f^2$, by differentiating this expression with respect to the factorization scale $\mu_f^2$ and  using the perturbatively calculated expressions for the  coefficient functions $C_{ij}$. 
In the physical-basis approach, on the other hand, we can first differentiate Eq.~\eqref{eq: structure function general notation} directly with respect to the physical scale $\log Q^2$, rather than the arbitrary factorization scale $\mu_f^2$. We then  invert the relation \eqref{eq: structure function general notation} to express the PDFs in terms of the structure functions as
\begin{equation}
\label{eq: pdfs in PB general notation}
    f_j(x,\mu_f^2) =  \sum _i C^{-1}_{ij}\left(\as(\mu_r^2), Q^2,\mu_f^2,\mu_r^2 \right)\otimes F_i(Q^2)\,.
\end{equation}
and substitute those expressions  back into  Eq.~\eqref{eq: structure function general notation}.
This yields  evolution equations directly for the dependence of physical observables on the physical scale $Q^2$, without reference to PDFs:
\begin{align}
\label{eq: structure function DGLAP in PB general notation}
   \frac{\dd F_{i}\xq}{\dd \log(Q^2)} &= \sum _j  
   \left[
   \frac{\dd }{\dd \log(Q^2)} C_{ij}\left(\as(\mu_r^2), Q^2,\mu_f^2,\mu_r^2 \right)
\right]
   \otimes 
      \sum _k C^{-1}_{kj}\left(\as(\mu_r^2), Q^2,\mu_f^2,\mu_r^2 \right)\otimes F_k(Q^2) \nonumber \\ &
   \equiv \sum_k P_{ik}(\as(\mu_r^2), Q^2)\otimes F_k(Q^2)\,.
\end{align}
Here the factorization scheme and scale dependence cancel within the physical basis evolution kernels $P_{ik}$. The physical-basis evolution equation is based on perturbative calculations, hence the renormalization scheme in the running coupling~$\as(\mu^2_r)$ remains as the only unphysical scale. The procedure can be continued to any perturbative order in $\as$.

In the conventional approach one parametrizes and fits unobservable, scheme-dependent quantities, the PDFs. This requires fitting PDFs to DIS and other collider data separately at every fixed perturbative order and factorization scheme. In a physical basis the initial values could be obtained by fitting DIS structure functions directly to data, at least in an ideal situation in which there would be enough data for all the required structure functions at a fixed value of $Q^2$. Even in the absence of a complete set of data for all observables, physical structure functions can be expected to be smooth, and satisfy unambiguous positivity constraints.
In contrast to the PDF-based approach, in a physical basis the initial condition for the evolution  always remains the same and, at least conceptually, independent of the factorization scale, scheme and perturbative order.
In any case, the physical-basis approach provides a more reliable and unambiguous interpretation of whether experimental data agree with the predictions of pQCD calculations, and a more rigorous quantification of the perturbative effects in increasing orders in $\as$.
In addition to DIS, the physical-basis approach can be applied to all PDF-dependent cross sections by simply replacing the PDFs with their physical-basis counterparts using Eq.~(\ref{eq: pdfs in PB general notation}). In practice this can mean e.g. expressing inclusive cross sections in proton-proton collisions directly in terms of DIS structure functions.

In practice, since we already know the splitting functions for PDFs, the derivation of the physical basis evolution equations is easiest to achieve by starting from an expression for the structure functions in terms of PDFs, using the DGLAP equations for PDFs, and then again expressing the PDFs in terms of structure functions. The scheme and scale independence of physical observables guarantees that in this procedure the scheme and scale dependence cancel between the coefficient functions $C_{ij}$ and the PDF splitting functions, in such a way that the structure function evolution kernels $P_{ik}$ remain scale and scheme independent. 
Since these kernels are scheme independent by construction, one can derive them by starting from known expressions given in any scheme.
In this manuscript we choose the $\msbar$ scheme and fix $\mu_f^2=\mu_r^2=Q^2$, for which NLO results for coefficient functions and PDF splitting functions are readily available.

The concept of a physical basis was introduced already in Ref.~\cite{Furmanski:1981cw} and has since been discussed in the literature by several authors~\cite{Catani:1996sc,Blumlein:2000wh,Hentschinski:2013zaa,Harland-Lang:2018bxd,Blumlein:2021lmf,vanNeerven:1999ca,RuizArriola:1998er}.
However, these previous studies of a physical basis are not applicable in a global analysis of perturbative QCD due to their specialised nature. In most of the previous studies, the observable basis consists of only two structure functions which are connected to either the quark singlet and non-singlet or to the quark singlet and the gluon PDF. The DGLAP evolution for the observables is usually expressed in terms of the physical anomalous dimensions, which are the physical basis equivalent for the Mellin moments of the splitting functions. In Ref.~\cite{vanNeerven:1999ca} the evolution equations have been derived in momentum space in next-to-next-to-leading order in $\as$, but in a basis consisting of only a non-singlet structure function.
In our previous work~\cite{Lappi:2023lmi}, we established the physical basis for six linearly independent DIS structure functions in the case where we have only three massless quark flavours in lowest order (LO) in~$\as$. In this manuscript we extend the work into next-to-leading order (NLO), still in a basis of six independent structure functions.  To our knowledge, our work at NLO in momentum space with six observables is so far the most extensive attempt to construct a physical basis.
This paper is organized as follows. In section~\ref{sec:DGLAPevolutioninphysicalbasis} we calculate the DGLAP evolution for the six-observable physical basis. The results of the numerical implementation of the physical basis are presented in section~\ref{sec:num results}, and finally, the work is concluded in section~\ref{sec:conclusions}.

\section{Evolution of a physical basis with six observables at NLO}
\label{sec:DGLAPevolutioninphysicalbasis}
In the previous work \cite{Lappi:2023lmi} we constructed a six-observable physical basis in the lowest order in~$\as$. The procedure of establishing the physical basis at NLO is similar as at LO, but more complicated due to the NLO corrections in the DIS structure functions and in the DGLAP splitting functions. As previously, here we only consider the massless 3-flavour basis for quarks, where $\nf=3$ and $s=\overline{s}$. The Cabibbo–Kobayashi–Maskawa mixing is also ignored here but can be straightforwardly incorporated in further studies.

For the NLO physical basis, we choose the DIS structure functions $\ft$, $\fk$, $\fkw$, $\ftcw$, and $\fl$ which we also used in the LO basis. However, instead of the structure function $\ftwminus$, we now use $\ftwdelta\equiv \ftwminus-\ftwplus$. This change is mostly done for reasons of numerical stability. Together with the structure function $\fk$, the structure function $\ftwdelta$ gives a better constraint for what corresponds to the valence quark PDFs, since both  $\fk$ and  $\ftwdelta$ are linear combinations of the valence quark distributions. This choice leads to a more stable numerical implementation at small values of $x$, where both $\fk$ and $\ftwdelta$ are small, while the sea quark-dominated $\ftwminus$ and $\ftwplus$ separately are numerically much larger. 
In order to have a compact notation, let us define the quark singlet as previously:
\begin{equation}
\label{eq:pure signlet}
   \Sigma \xq \equiv  \sum_q^{\nf} \left(q\xq+\overline{q}\xq \right)\,,
\end{equation}
and the singlet weighted with the electric charges (charge-weighted singlet) as
\begin{equation}
\label{eq:signlet}
   \Xi\xq \equiv  \seq \left(q\xq+\overline{q}\xq \right)\,.
\end{equation}
The six DIS structure functions, chosen to span the physical basis, are written in the $\msbar$ scheme at NLO as
\begin{align}
    \label{eq:F2fullB}
    \ft (x) & = x \Xi(x) + \As x\left\{ \cftqnlo\otimes \Xi +2\nf\seqav\cftgnlo\otimes g \right\} \,, \\
    \label{eq:F3fullB}
    \fk (x) & = 2\sum_{q}(L^2_q-R_q^2) \left[ q(x) - \overline{q}(x) \right]  + \As 2\cfkqnlo\otimes\sum_{q}(L^2_q-R_q^2) \left[ q - \overline{q} \right]  \,, \\
     \label{eq:F2WdeltafullB}
    \ftwdelta (x)  &=  2x \left[ u(x)- \overline{u}(x) -(d(x)-\overline{d}(x)) \right] + \As 2x \cftqnlo\otimes\left[ u-\overline{u}-(d-\overline{d}) \right]    \,, \\
    \iffalse
    \label{eq:F2WminusfullB}
    \ftwminus (x) & = 2x \left[ u(x)+\overline{d}(x)+\overline{s}(x) \right] + \As 2x\left\{ \cftqnlo\otimes\left[ u+\overline{d}+\overline{s} \right]  +\nf \cftgnlo\otimes g \right\}  \,, \\
    \label{eq:F2WplusfullB}
    \ftwplus (x) & = 2x \left[ \overline{u}(x)+d(x)+s (x) \right] + \As 2x\left\{ \cftqnlo\otimes\left[ \overline{u}+d+s  \right]  +\nf \cftgnlo\otimes g \right\}  \,, \\
    \fi
    \label{eq:F3WfullB}
    \fkw (x) & = 2 \left[ u(x)-\overline{d}(x)-\overline{s}(x) \right]  + \As 2\cfkqnlo\otimes \left[ u-\overline{d}-\overline{s} \right] \,, \\
    \label{eq:F2CWfullB}
    \ftcw (x) & = 2x\overline{s}(x)  + \As 2x\left\{ \cftqnlo\otimes\overline{s} +\cftgnlo\otimes g \right\}   \,, \\
    \label{eq:FLfullB}
    \fl(x) & = \As x\left\{ \cflqlo\otimes \Xi  +2\nf\seqav\cflglo\otimes g\right\} \nonumber
    \\ &+\left(\As\right)^2 x\Bigg\{ \seqav \cflpsnlo\otimes\Sigma   + \cflnsnlo\otimes \Xi   +2\nf\seqav\cflgnlo\otimes g\Bigg\}\,, 
\end{align}
where $\seqav$ is the averaged electric charge
\begin{equation}
    \label{eq:eq2av}
   \seqav\equiv \frac{1}{\nf}\sum_q e_q^2 \,,
\end{equation}
and the symbol $\otimes$ denotes the convolution 
\begin{equation}
\label{eq:convolution2}
f\otimes g \equiv \int_{x}^{1} \frac{\dd z}{z}f(z)g\left(\frac{x}{z}\right) \,. 
\end{equation}
In Eq.~\eqref{eq:F3fullB} we have used $L_q = T_q^3-2e_q\sin^2 \theta_W$ and $R_q =-2e_q\sin^2 \theta_W$, where $T_q^3$ is the third component of the weak isospin and $\theta_W$ denotes the Weinberg angle.  For completeness, we list all the coefficient functions in our notation in Appendix~\ref{appendix C and P}.  The $\msbar$ coefficient functions in the definition of $\fl$ are taken from Ref.~\cite{Moch:2004xu}.
The rest of the NLO $\msbar$ coefficient functions, corresponding to the 1-loop corrections, are consistent with Ref.~\cite{Ellis:1996mzs}. Note that we have used a different normalization convention for the coefficient functions than  the original sources. The running coupling constant in Eqs.~\eqref{eq:F2fullB}--\eqref{eq:FLfullB} follows the renormalization group equation
\begin{equation}
\label{eq: renormalization group eq}
 \frac{\dd\as(\mu^2_r)}{\dd \log(\mu^2_r)} = - \frac{b_0}{2\pi} \alpha^2_s(\mu^2_r)-\frac{b_1}{(2\pi)^2} \alpha^3_s(\mu^2_r)+\mathcal{O}(\as^4) \,, 
\end{equation}
where \begin{equation}
b_0 = \frac{11C_A-4T_Rn_f}{6} \,, b_1 = \frac{34 C_A^2-12\cf\TR\nf-20C_A\TR\nf}{12} \,, \ C_A = 3 \,, \ T_R = 1/2 \,.
\end{equation} 
Note that we take here the point of view that NLO means the second \emph{nonzero} order in perturbation theory, and include $\fl$ up to order $\as^2$. This is a natural approach also from the point of view of the dipole picture of DIS at small $x$, where $\fl$ and $\ft$ are parametrically of the same order in $\as$~\cite{Kovchegov:2012mbw}. 

Establishing a physical basis requires first inverting the PDFs, which means expressing them in terms of the structure functions. 
If we consider all the physical observables as fixed (i.e. not as expansions in $\as$), the PDFs in turn can be considered as a series in the coupling: 
\begin{align}
    \label{eq:PDF P expansion}
   f_i\xq & = \sum _n \left(\As\right)^n f_i^{(n)}\xq  \,,
\end{align}
where $n\in\{0,1,2,...\}$.
Now we can utilise the LO physical-basis results~\cite{Lappi:2023lmi}, where we were able to invert the relation from PDFs to observables at leading order, to construct an iterative inversion procedure organized in powers of $\as$. We will denote the PDF approach distributions with lowercase Latin letters  $g,u,\bar{u}, \dots$. Their physical basis counterparts, which are considered to be calculated from the structure functions in the perturbative expansion \eqref{eq:PDF P expansion}, are denoted by the corresponding uppercase letters $G,U,\bar{U}, \dots$.

As a starting point, the expressions for the LO charge-weighted singlet and gluon are 
\begin{align}
    \label{eq:singlet PB LO}
     \singletlo\xq &= \wft \xq \,,\\
    \label{eq:gluon PB LO}
     \Glo\xq &= \frac{1}{8\TR\nf\seqav}\left\{2\cf \left(x\frac{\dd}{\dd x}-2 \right)\wft\xq+\hat{P}(x)\wfl\xq \right\} \,,
\end{align}
where we have defined a differential operator
\begin{equation}
\label{eq:Phat}
    \hat{P}(x) \equiv x^2 \frac{\dd^2}{\dd x^2}-2x\frac{\dd}{\dd x}+2 \,.
\end{equation}
For brevity we have introduced the following notation for the structure functions (with $\fl$ scaled by $\as$) and their derivatives
\begin{align}
\wft\xq & \equiv \frac{\ft(x, Q^2)}{x} \,, \\
\label{eq:wfl}
\wfl\xq &\equiv  \frac{2\pi}{\as(\mu_r^2)} \frac{\fl(x, Q^2)}{x} \,, \\
\label{eq:f2lprime}
\widetilde{F'}_{2,L}\xq &\equiv  x\frac{\dd}{\dd{x}}\widetilde{F}_{2,L}(x, Q^2) \,, \\
\label{eq:wflpp}
\wflpp\xq &\equiv x^2\frac{\dd[2]}{\dd{x^2}}\wfl(x, Q^2) \,.
\end{align}

The NLO physical-basis counterpart for the charge-weighted quark singlet is obtained by moving the NLO component from Eq.~\eqref{eq:F2fullB} to the right hand side
\begin{align}
    \label{eq:singlet inversion}
    \Xi\xq &= \wft - \As \left\{ \cftqnlo\otimes\Xi(Q^2) +2\nf\seqav\cftgnlo\otimes g(Q^2) \right\} \,,
\end{align}
and then inserting the LO expressions for the singlet and the gluon from Eqs.~\eqref{eq:singlet PB LO} and \eqref{eq:gluon PB LO}
\begin{align}
    \label{eq:singlet PB NLO}
    \Xi\xq &= \wft - \As \left\{ \cftqnlo\otimes\wft +2\nf\seqav\cftgnlo\otimes \Glo \right\} +\mathcal{O}(\as^2) \,.
\end{align}
Note that while we have, for brevity, not written it out explicitly, the $\Glo$ in this equation is understood to be taken as the r.h.s. of Eq.~\eqref{eq:gluon PB LO}, i.e. an expression in terms of structure functions $\ft$ and $\fl$ and their derivatives. We have thus expressed the charge-weighted singlet distribution in terms of structure functions and their derivatives. 
Using a similar iterative procedure, all the quark PDFs can be expressed in the physical basis. We have listed all the inverted PDFs and the quark singlet at NLO in Appendix~\ref{appendix PDFs the in physical basis}.  This iterative procedure would be straightforward  to continue to higher orders in $\as$, with higher derivatives of the structure functions appearing at each order.

An equally straightforward technique can not be applied to invert the gluon PDF due to the convolutions present in the LO component of the structure function $\fl$. However, one can define the inverse of the LO convolution with the gluon PDF, in the definition of $\fl$ in Eq.~\eqref{eq:FLfullB}, by using the differential operator defined in Eq.~\eqref{eq:Phat}:
\begin{equation}
    g\xq = \frac{1}{4\TR} \hat{P}(x) \left[ \cflglo\otimes g(Q^2) \right]\,. \label{eq:invertG}
\end{equation}
Now we can solve the LO contribution of the gluon distribution from the expression for $\fl$ in Eq.~\eqref{eq:FLfullB} 
\begin{align}
\label{eq:cflglo conv g}
    \cflglo\otimes g(Q^2)  = \wfl\xq -\cflqlo \otimes\Xi(Q^2)  - \As \Big[ \seqav\cflpsnlo\otimes \Sigma (Q^2)+\cflnsnlo\otimes\Xi(Q^2)  + 2\nf\seqav\cflgnlo\otimes g(Q^2)  \Big]\,,
\end{align}
and extract the gluon distribution by operating with the inverse of the LO  coefficient function $\cflglo$, i.e. the differential operator $\hat{P}$. The PDFs on the right hand side of Eq.~\eqref{eq:cflglo conv g} are   known in terms of the physical basis structure functions, noting that in the higher order term $\sim \as  G(Q^2)$ one can use the LO gluon $\Glo(Q^2)$.
We thus arrive with the physical-basis counterpart for the NLO gluon 
\begin{align}
\label{eq:gluon PB full}
    G\xq = &\frac{1}{8\TR\nf\seqav} \hat{P}(x)  \Bigg\{ \wfl\xq -\cflqlo \otimes\Xi(Q^2) \nonumber  \\& - \As \Big[ \seqav\cflpsnlo\otimes \Sigma (Q^2)+\cflnsnlo\otimes\Xi(Q^2)  + 2\nf\seqav\cflgnlo\otimes G(Q^2)  \Big]  \Bigg\}\nonumber  \\
    =& \frac{1}{8\TR\nf\seqav}\Bigg\{2\cf \left(x\frac{\dd}{\dd x}-2 \right)\wft\xq+\hat{P}(x)\wfl\xq \Bigg\}  \nonumber  \\& - \As  \frac{1}{8\TR\nf\seqav}\hat{P}(x) \Bigg\{ \seqav\cflpsnlo\otimes\PSlo(Q^2)  +\cflnsnlo\otimes\wft(Q^2) \nonumber   \\&+2\nf\seqav\cflgnlo\otimes \Glo(Q^2)  + \cflqlo \otimes \singletnlo (Q^2) \Bigg\} +\mathcal{O}(\as^2)\,.
\end{align}
The complete expression, with $\singletlo$, $\Glo$, and $\PSlo$ explicitly written in the physical basis, is given in Appendix~\ref{appendix PDFs the in physical basis} in Eq.~\eqref{eq:NLO gluon PB}. From the view point of deriving the evolution equations it is here easiest to keep the differential operator $\hat{P}$ outside of the convolutions -- the gluon PDF in physical basis will be convoluted with other functions in the DGLAP evolution equations, allowing $\hat{P}$ to be removed by partial integration. However, as we will discuss briefly later on, in practical applications to processes others than the ones included in the physical basis will require separately the numerical values for the LO and NLO parts of the perturbative expansions of PDFs, e.g. $G^{(0)}$ and $G^{(1)}$ given in the equation above. In this case one will need to execute the derivatives on the structure functions and convolutions appearing in~Eq.\eqref{eq:gluon PB full} directly either numerically or partly analytically. In doing this, attention has to be paid to the boundary conditions, since some of the NLO coefficient functions involve plus distributions as well as delta functions. 

The iterative inversion of the PDFs can be carried out in a similar manner to higher orders of~$\as$. Here, at NLO, we have two nested $\hat{P}$ operators, with the second one inside $\Glo$, see Eq.~\eqref{eq:gluon PB LO}, and thus not explicit in Eq.~\eqref{eq:gluon PB full}. This indicates that at next-to-NLO the highest number of nested differential operators would be three. An accurate handling of the resulting multiple derivatives at higher powers of~$\as$ is something that can potentially pose a challenge in the physical-basis approach. The derivatives are merely a technical issue in the numerical implementation. One has to utilize a numerical approach which is able to reliably describe derivatives of all orders in the equations. The appearance of derivatives is a qualitatively distinct feature of the physical-basis approach in contrast to the fully PDF-based picture.

As discussed in Sec.~\ref{sec:intro}, while in principle we could start deriving the DGLAP equations from the logarithmic parts of the coefficient functions, in practice it is easiest to just use the known PDF splitting functions at the scale $\mu_f^2=Q^2$.  Thus
we obtain the DGLAP evolution equations for the structure functions by taking the $\log(Q^2)$ derivative from the definitions in Eqs.~\eqref{eq:F2fullB} -- \eqref{eq:FLfullB}, where the PDFs and the running coupling $\as(Q^2)$ are affected by the derivative. 
The NLO DGLAP evolution for the PDFs is given by \cite{Ellis:1996mzs}
\begin{align}
    \label{eq: quark DGLAP}
    \frac{\dd q_i\xq}{\dd \log(Q^2)} &= \As  \Bigg\{\Pqqlo\otimes q_i(Q^2) + \Pqglo\otimes g(Q^2) \nonumber \\ &+ \As \Bigg[\PVqqnlo \otimes q_i(Q^2) +\PVqqbarnlo \otimes \overline{q}_i(Q^2)+\Psingletnlo\otimes \sum_j (q_j(Q^2)+\overline{q}_j(Q^2))  +\Pqgnlo\otimes g (Q^2)\Bigg]  \Bigg\}  \,, \\
    \frac{\dd \overline{q}_i\xq}{\dd \log(Q^2)} &=  \As  \Bigg\{\Pqqlo\otimes \overline{q}_i(Q^2) + \Pqglo\otimes g(Q^2) \nonumber \\ &+ \As \Bigg[\PVqqbarnlo \otimes q_i(Q^2) +\PVqqnlo \otimes \overline{q}_i(Q^2)+\Psingletnlo\otimes \sum_j (q_j(Q^2)+\overline{q}_j(Q^2))  +\Pqgnlo\otimes g (Q^2)\Bigg]  \Bigg\} \,, 
    \label{eq: antiquark DGLAP}\\ 
    \frac{\dd g\xq}{\dd \log(Q^2)} &=   \As  \Bigg\{\left[\Pgqlo+\As\Pgqnlo\right]\otimes \sum_j (q_j(Q^2)+\overline{q}_j(Q^2)) + \left[ \Pgglo+\As \Pggnlo\right] \otimes g(Q^2) \Bigg\}\,.
    \label{eq: gluon DGLAP}
\end{align}
The splitting functions are listed in Appendix~\ref{appendix C and P}. 

By combining the renormalization group equation \eqref{eq: renormalization group eq}, the DGLAP evolution equations for the PDFs in Eqs.~\eqref{eq: quark DGLAP}--\eqref{eq: gluon DGLAP}, and the PDFs the in physical basis in Eqs.~\eqref{eq:dLO}--\eqref{eq:sLO} and \eqref{eq:gluon PB full}, we derive the physical-basis DGLAP evolution equations for the structure functions:
\begin{align}
    \label{eq: F2 DGLAP}
\frac{\dd \ft\xq}{\dd \log(Q^2)} = & \As x \left\{ \Pqqlo\otimes\wft +2\nf\seqav\Pqglo \otimes\Glo \right\} +\left( \As\right)^2x\Bigg\{ \Pplusnlo \otimes\wft +2\nf\seqav\Psingletnlo \otimes\PSlo +2\nf\seqav\Pqgnlo\otimes\Glo  
\nonumber \\& 
- 2\nf\seqav\Pqqlo\otimes\cftgnlo\otimes\Glo +2\nf\seqav\Pqglo\otimes\Gnlo 
+2\nf\seqav\cftgnlo\otimes\left[\Pgqlo\otimes\PSlo + \Pgglo\otimes\Glo \right]
\nonumber \\&
+2\nf\seqav\cftqnlo\otimes\Pqglo\otimes\Glo
-b_0\left[ \cftqnlo\otimes\wft+2\nf\seqav\cftgnlo\otimes\Glo \right]
\Bigg\}     +\mathcal{O}(\as^3) \,, \\
  \label{eq: F3 DGLAP}
 \frac{\dd \fk\xq}{\dd \log(Q^2)} = &\As \Pqqlo\otimes\fk  +\left( \As\right)^2\left\{ \Pminusnlo \otimes\fk -b_0\cfkqnlo\otimes\fk
\right\}     +\mathcal{O}(\as^3) \,, \\
  \label{eq: F2deltaW DGLAP}
\frac{\dd \ftwdelta\xq}{\dd \log(Q^2)} =& \As x\Pqqlo\otimes\wftwdelta +\left( \As\right)^2 x \left\{ \Pminusnlo \otimes\wftwdelta -b_0\cftqnlo\otimes\wftwdelta
\right\}     +\mathcal{O}(\as^3) \,, \\
\label{eq: F3W DGLAP}
\frac{\dd \fkw\xq}{\dd \log(Q^2)} = &\As \left\{ \Pqqlo\otimes\fkw -2\Pqglo \otimes\Glo \right\} 
+\left( \As\right)^2\Bigg\{ \PVqqnlo \otimes\fkw \nonumber \\& +\frac{1}{A_d+A_u}\PVqqbarnlo\otimes\left( -2\fk -(A_d-A_u)\wftwdelta+(A_d+A_u)\fkw \right)
-2\Psingletnlo \otimes\PSlo 
\nonumber \\&
-2\Pqgnlo\otimes\Glo  -2\Pqglo\otimes\Gnlo 
-2\cfkqnlo\otimes\Pqglo\otimes\Glo
-b_0\cfkqnlo\otimes\fkw
\Bigg\}     +\mathcal{O}(\as^3) \,, \\
  \label{eq: F2CW DGLAP}
\frac{\dd \ftcw\xq}{\dd \log(Q^2)} &= \As x \left\{ \Pqqlo\otimes\wftcw +2\Pqglo \otimes\Glo \right\} 
+\left( \As\right)^2 x \Bigg\{ \Pplusnlo \otimes\wftcw +2\Psingletnlo \otimes\PSlo \nonumber \\& +2\Pqgnlo\otimes\Glo -2\Pqqlo\otimes\cftgnlo\otimes\Glo +2\Pqglo\otimes\Gnlo 
+2\cftgnlo\otimes\left[\Pqglo\otimes\PSlo + \Pgglo\otimes\Glo \right]\nonumber \\&+2\cftqnlo\otimes\Pqglo\otimes\Glo
-b_0 \left[ \cftqnlo\otimes\wftcw+2\cftgnlo\otimes\Glo \right]
\Bigg\}     +\mathcal{O}(\as^3) \,, 
\end{align}
\begin{align}
  \label{eq: FL DGLAP}
&\frac{\dd }{\dd \log(Q^2)} \left(\frac{\fl \xq}{\As}\right) 
\nonumber  = \\ & 
\quad \quad 
 \As x \left\{  \cflqlo\otimes\left[\Pqqlo\otimes\wft+2\nf\seqav\otimes\Pqglo\otimes\Glo \right]+2\nf\seqav\cflglo\otimes\left[\Pgqlo\otimes\PSlo+\Pgglo\otimes\Glo \right] \right\} 
\nonumber \\& 
\quad \quad 
+\left( \As\right)^2 x\Bigg\{ \cflqlo\otimes\Big[ \Pplusnlo\otimes\wft +2\nf\seqav\Psingletnlo\otimes\PSlo+2\nf\seqav\Pqgnlo\otimes\Glo 
\nonumber \\&
\quad \quad 
\quad \quad 
-\Pqqlo\otimes\left( \cftqnlo\otimes\wft+2\nf\seqav\cftgnlo\otimes\Glo \right) 
 +2\nf\seqav\Pqglo\otimes\Gnlo \Big] 
 \nonumber \\&
 \quad \quad 
+2\nf\seqav\cflglo\otimes\left[ \Pgqnlo\otimes\PSlo+\Pggnlo\otimes\Glo+\Pgqlo\otimes\PSnlo+\Pgglo\otimes\Gnlo \right]
 \nonumber \\& 
 \quad \quad 
+\seqav\cflpsnlo\otimes\left[\Pqqlo\otimes\PSlo+2\nf\Pqglo\otimes\Glo \right] +\cflnsnlo\otimes\left[\Pqqlo\otimes\wft+2\nf\seqav\Pqglo\otimes\Glo \right] 
\nonumber \\& 
\quad \quad 
+2\nf\seqav\cflgnlo\otimes\left[\Pgqlo\otimes\PSlo+\Pgglo\otimes\Glo \right] -b_0\left[\seqav\cflpsnlo\otimes\PSlo+\cflnsnlo\otimes\wft+2\nf\seqav\cflgnlo\otimes\Glo \right]
\Bigg\}     +\mathcal{O}(\as^3) \,.
\end{align}
Here we have adopted the notation
\begin{align}
    \label{eq: Pqq+}
\Pplusnlo &\equiv \PVqqnlo + \PVqqbarnlo \,, \\ 
    \label{eq: Pqq-}
\Pminusnlo  &\equiv \PVqqnlo - \PVqqbarnlo\,.
\end{align}
For brevity, we have not substituted the explicit expressions for $\PSlo$, $\PSnlo$, $\Glo$, and $\Gnlo$, which can be taken from Eqs.~\eqref{eq:pure singlet LO phys basis}, \eqref{eq:pure singlet NLO phys basis}, \eqref{eq:gluon PB LO}, and \eqref{eq:gluon PB full}.

In the expressions above the inverted gluon $G$ appears in convolutions with the LO splitting functions and coefficient functions. Thus, the differential operator $\hat{P}$ in the inverted gluon can be removed with partial integration, the Leibniz integration rule, and careful management of the boundary conditions. 
Following the partial integration, the DGLAP evolution equations for the structure functions can be finally expressed as the convolutions of the evolution kernels with the structure functions
\begin{equation}
    \label{eq: kernel expression DGLAP }
   \frac{\dd F_i\xq}{\dd \log(Q^2)}= \sum_j P_{F_i F_j}\otimes F_j+\sum_j P_{F_i F'_j}\otimes F'_j+\sum_j P_{F_i F''_j}\otimes F''_j \,,
\end{equation}
where $F_i, F_j \in \{\wft, \fk, \wftwdelta, \fkw, \wftcw, \wfl \}$. While the solutions to the evolution equations above are unique, the evolution kernels introduced above can be formulated in different ways. By continuing partial integration one could transfer terms between kernels $P_{F_i F_j''}$, $P_{F_i F_j'}$, and $P_{F_i F_j}$. Also, the derivatives of the structure functions could be absorbed into the evolution kernels by expressing them with the delta function derivatives. Derivatives acting on delta functions would, however, in practice need to be partially integrated away for a numerical implementation, so we prefer not to present the kernels in such a formal way. In this paper we present one possible way to write the evolution kernels. With our choice,  all the non-zero kernels needed in the physical basis DGLAP evolution equations are:
\begin{align}
    \label{eq: kernels F2 DGLAP}
\frac{\dd \ft\xq}{\dd \log(Q^2)} &= x\Big\{\sum_i P_{\ft F_i}\otimes F_i + \sum_j P_{\ft F'_j}\otimes F'_j + P_{\ft \wflpp}\otimes\wflpp \Big\}\,, \\
    \label{eq: kernels F3 DGLAP}
\frac{\dd \fk\xq}{\dd \log(Q^2)} &= P_{\fk \fk} \otimes\fk \,, \\
    \label{eq: kernels deltaF2 DGLAP}
\frac{\dd \ftwdelta\xq}{\dd \log(Q^2)} &= x P_{\ftwdelta \wftwdelta}\otimes\wftwdelta \,, \\
    \label{eq: kernels F3W DGLAP}
\frac{\dd \fkw\xq}{\dd \log(Q^2)} &=  \sum_i P_{\fkw F_i}\otimes F_i + \sum_j P_{\fkw F'_j}\otimes F'_j + P_{\fkw \wflpp}\otimes\wflpp  \,, \\
    \label{eq: kernels F2CW DGLAP}
\frac{\dd \ftcw\xq}{\dd \log(Q^2)} &= x\Big\{P_{\ftcw \wftcw}\otimes\wftcw+\sum_i P_{\ftcw F_i}\otimes F_i + \sum_j P_{\ftcw F'_j}\otimes F'_j + P_{\ftcw \wflpp}\otimes\wflpp \Big\} \,, \\
    \label{eq: kernels FL DGLAP}
\frac{\dd }{\dd \log(Q^2)} \left(\frac{\fl \xq}{\As}\right) &=  x\Big\{\sum_i P_{\fl F_i}\otimes F_i + \sum_j P_{\fl F'_j}\otimes F'_j + P_{\fl \wflpp}\otimes\wflpp \Big\} \,, 
\end{align}
where $F_i\in \{\wft, \fk, \wftwdelta, \fkw, \wfl\}$ and $F'_j\in \{\wftp,  \wflp\}$.  We have listed these evolution kernels in Appendix~\ref{appendix Evolution kernels}. These expressions are relatively straightforward to convert into a numerical form, especially compared to the expressions that would arise if the structure function derivatives would be absorbed into the evolution kernels. \footnote{We compared our results, in the approximation where we have only the structure functions $\ft$ and $\fl$, to the corresponding Mellin space results in Ref.~\cite{Hentschinski:2013zaa}. It appears that there may be an inconsistency in the evolution kernels presented in Ref.~\cite{Hentschinski:2013zaa}, between the splitting functions $P_{qg}$ and $P_{gq}$, which seems to affect the numerical results.}

\begin{figure}[htb]
  \includegraphics[width=.5\linewidth]{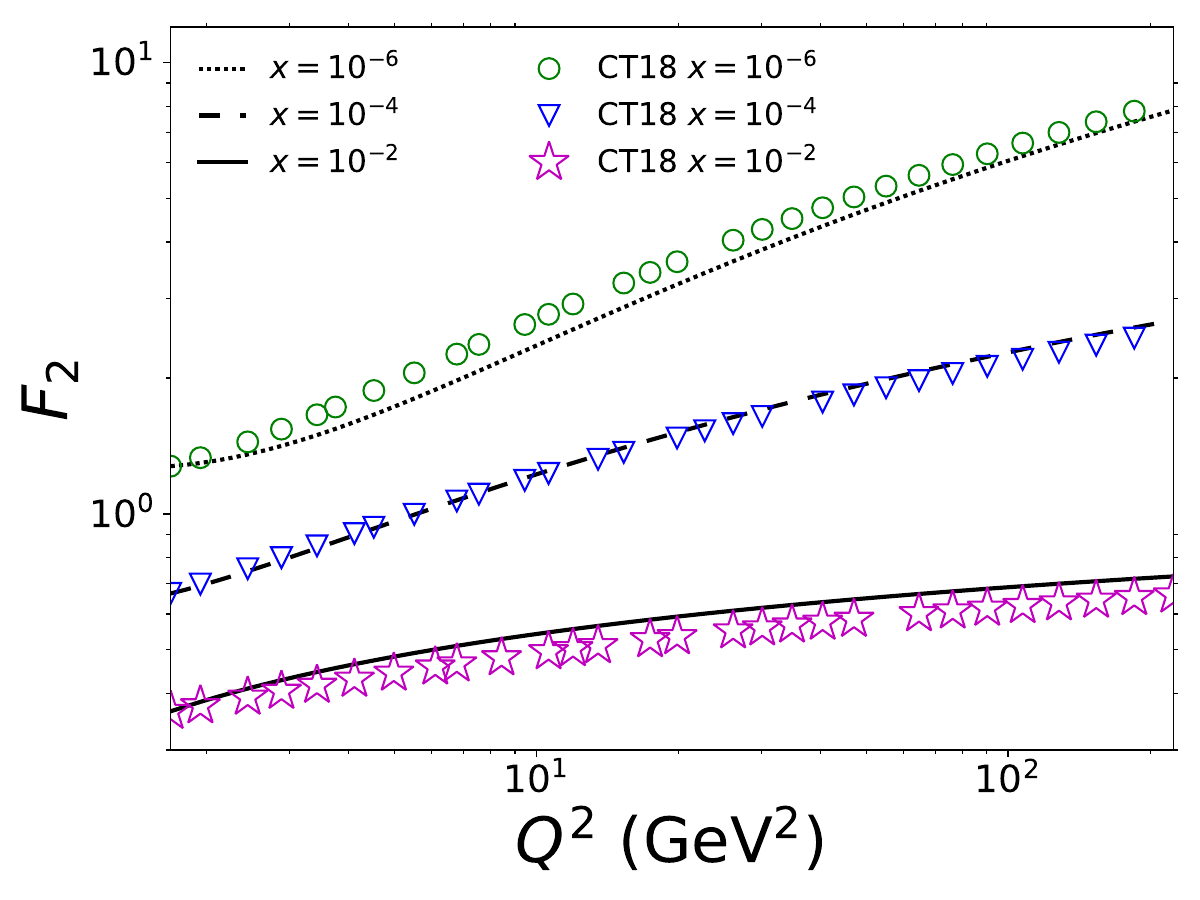}\hfill
  \includegraphics[width=.5\linewidth]{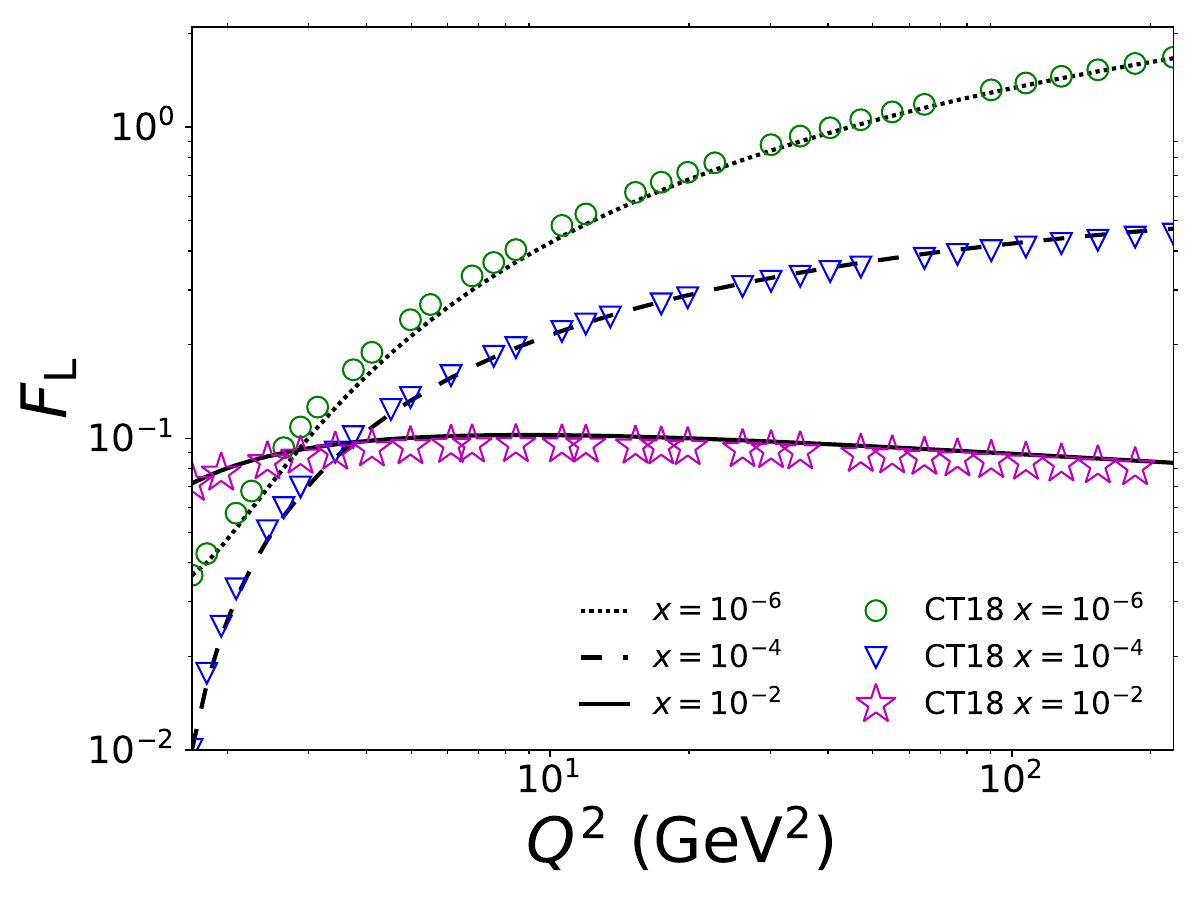}
  \includegraphics[width=.5\linewidth]{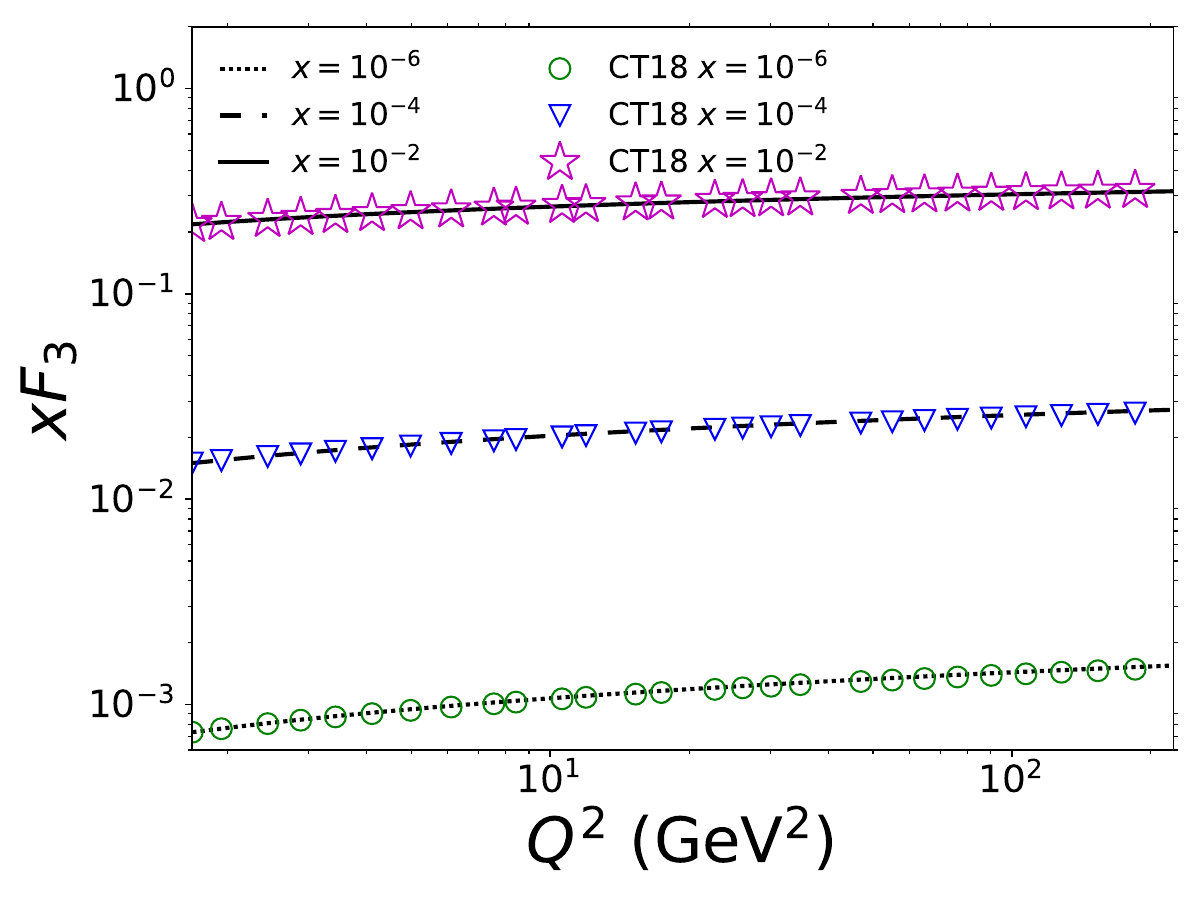}\hfill
  \includegraphics[width=.5\linewidth]{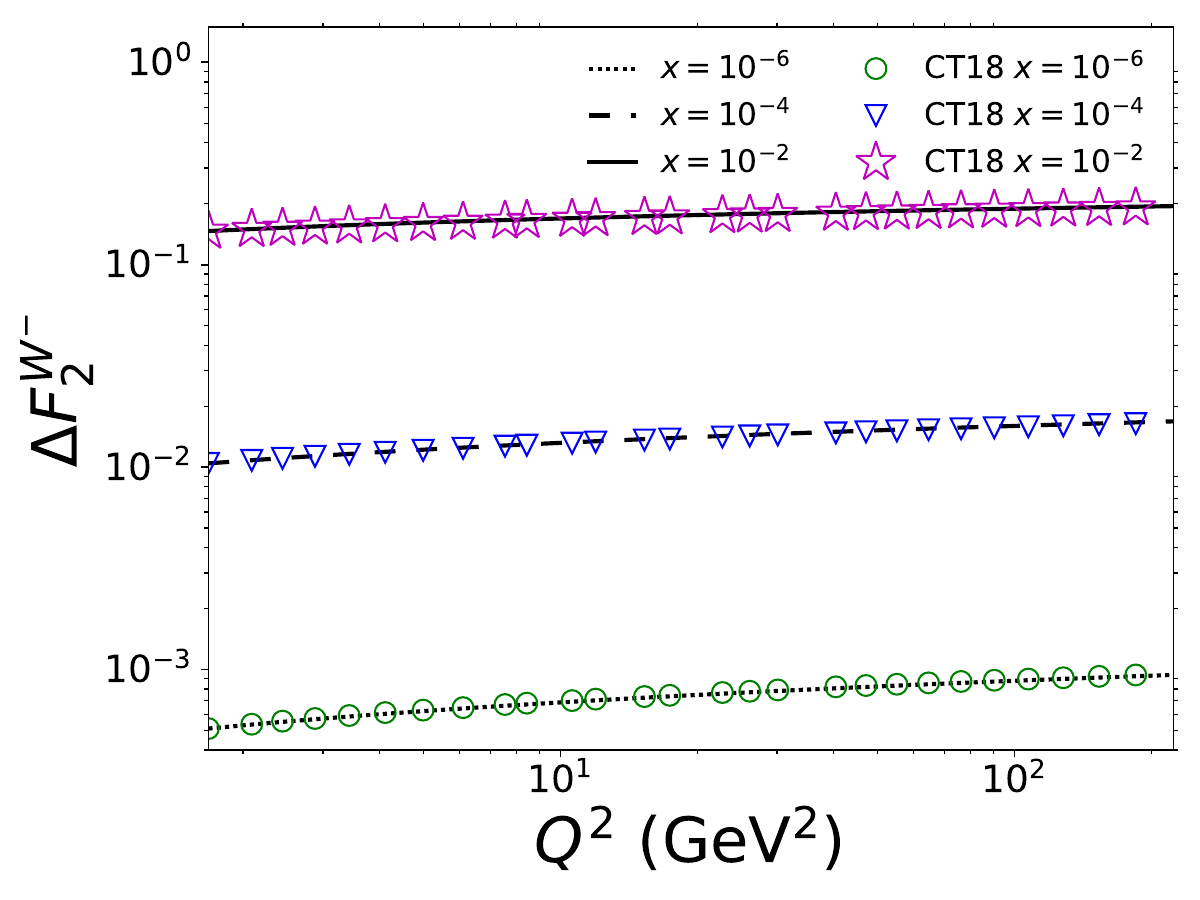}
  \includegraphics[width=.5\linewidth]{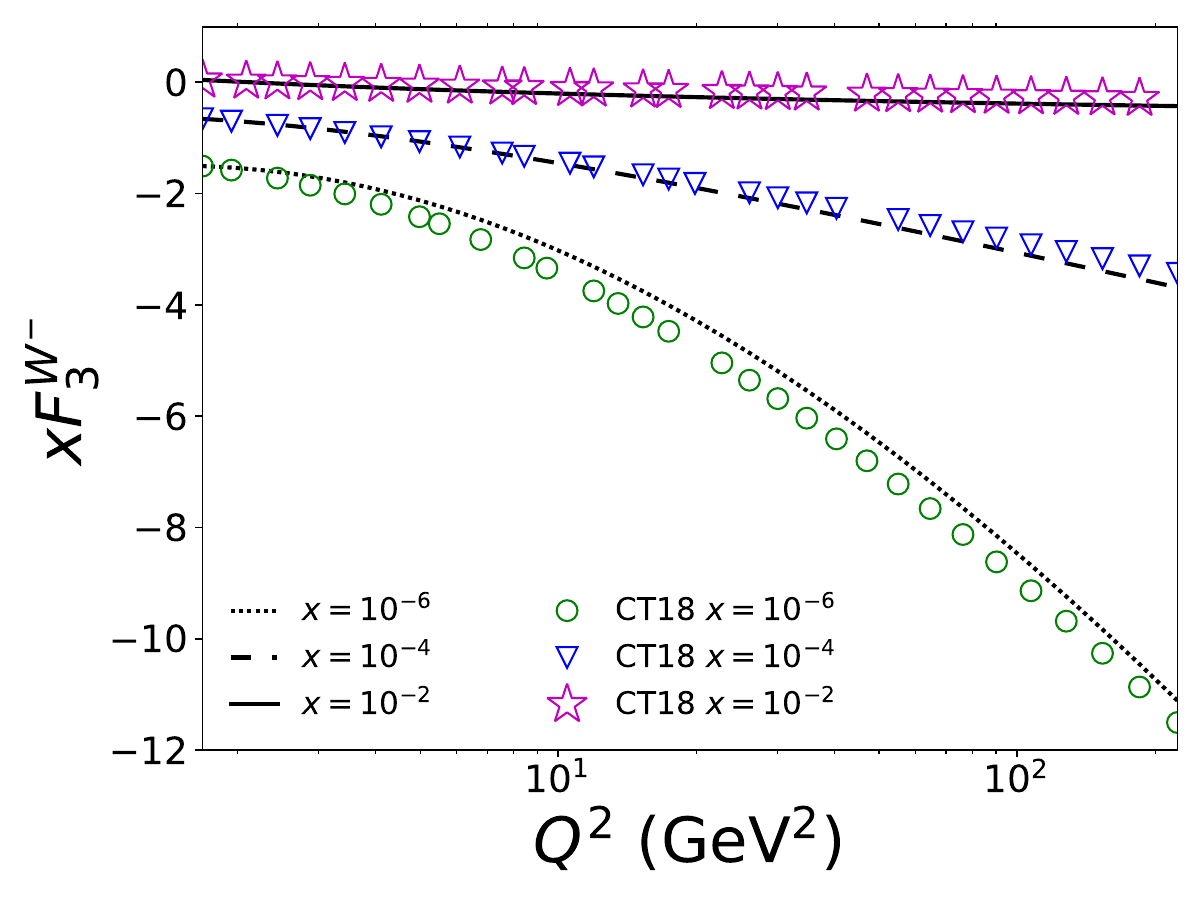}\hfill
  \includegraphics[width=.5\linewidth]{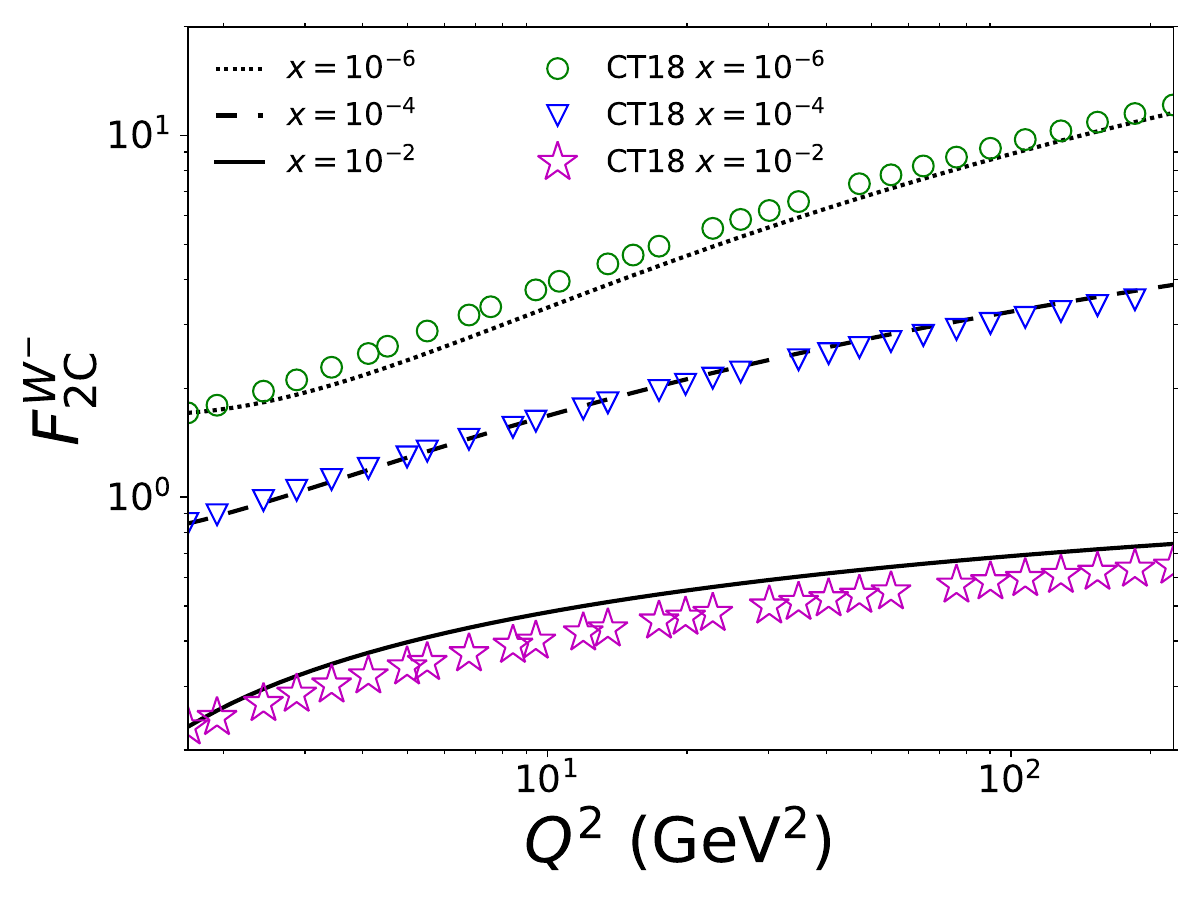}
  
  \caption{
  The $Q^2$ dependence of $\ft$, $\fl$, $\fk$, $\ftwdelta$, $\fkw$, and $\ftcw$ using the physical-basis approach (curves) compared with the usual PDF-based approach (markers).}
\label{fig:Q2resultsFullBasis}
\end{figure}

\begin{figure}[htb]
  \includegraphics[width=.5\linewidth]{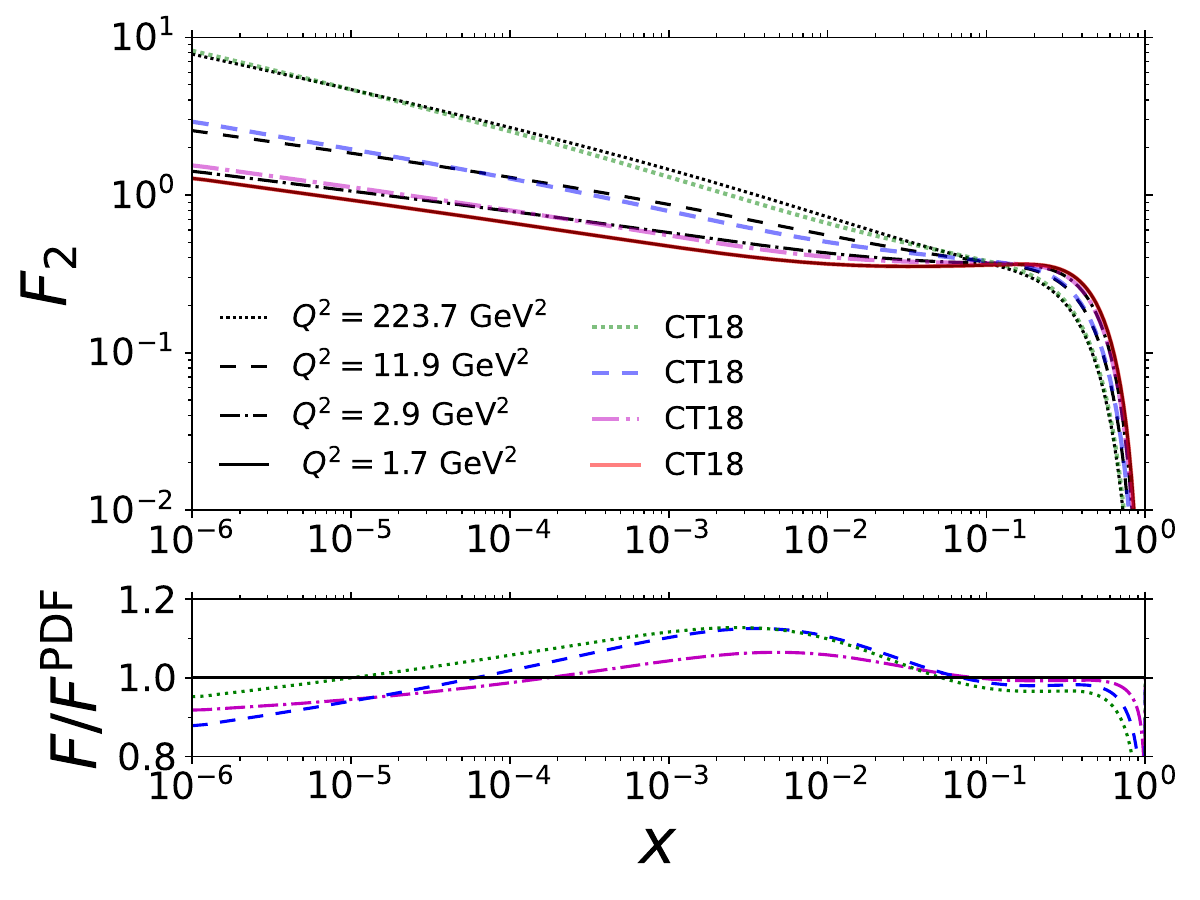}\hfill
  \includegraphics[width=.5\linewidth]{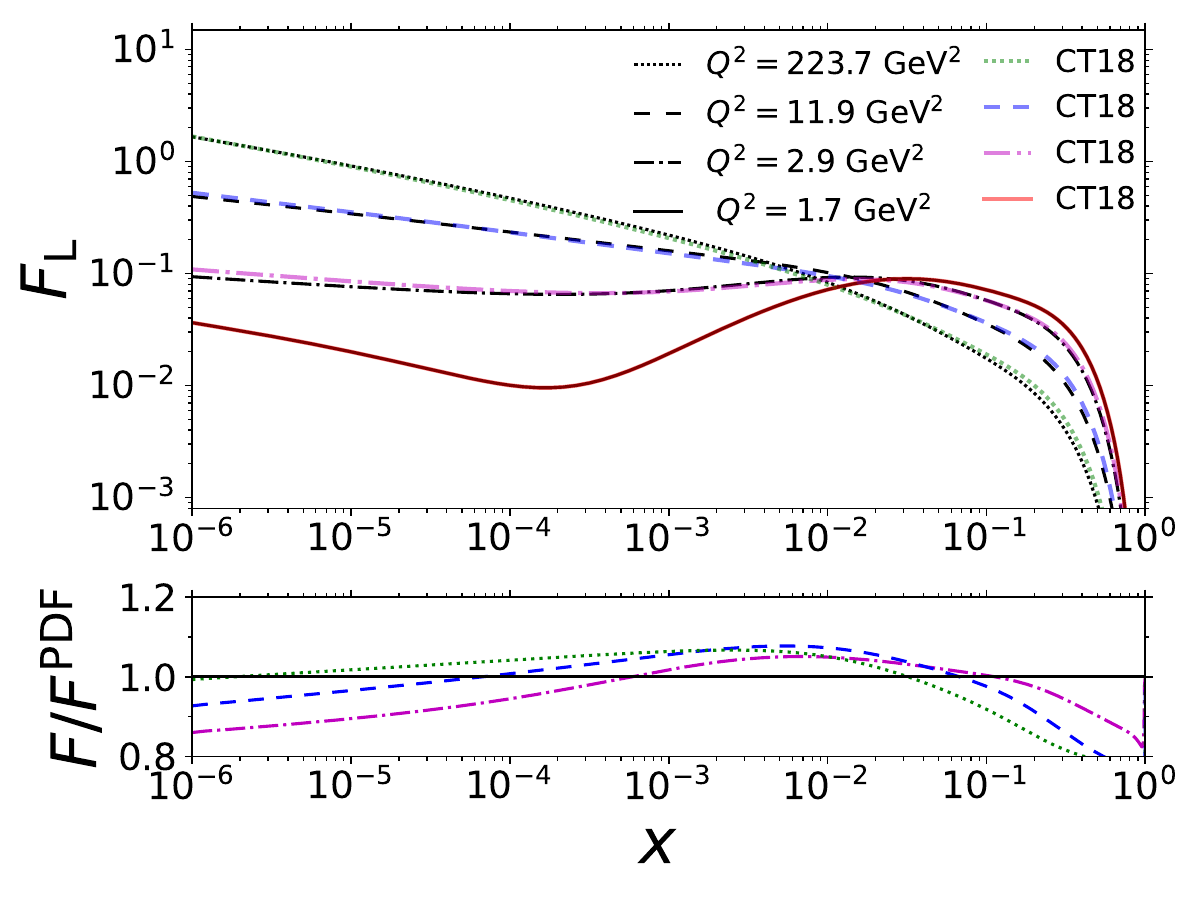}
  \includegraphics[width=.5\linewidth]{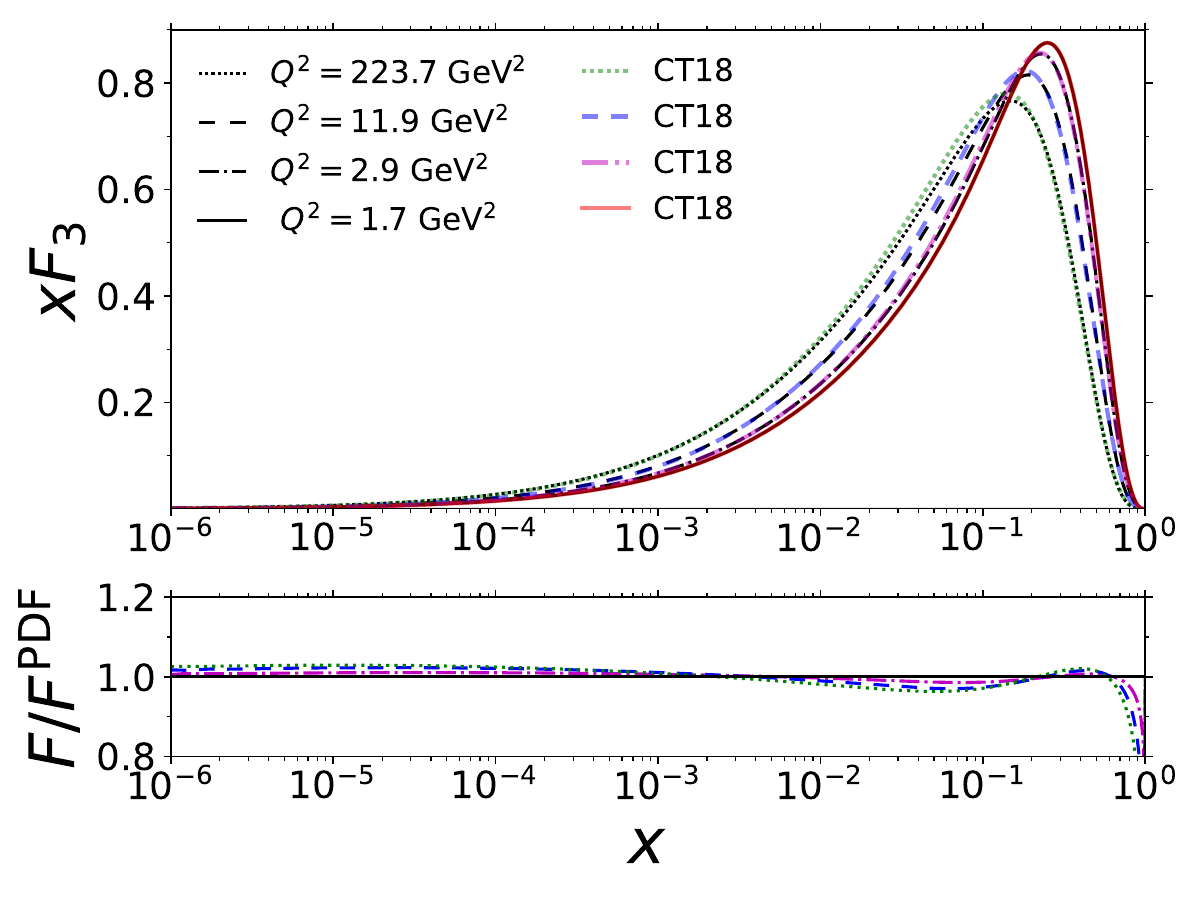}\hfill
  \includegraphics[width=.5\linewidth]{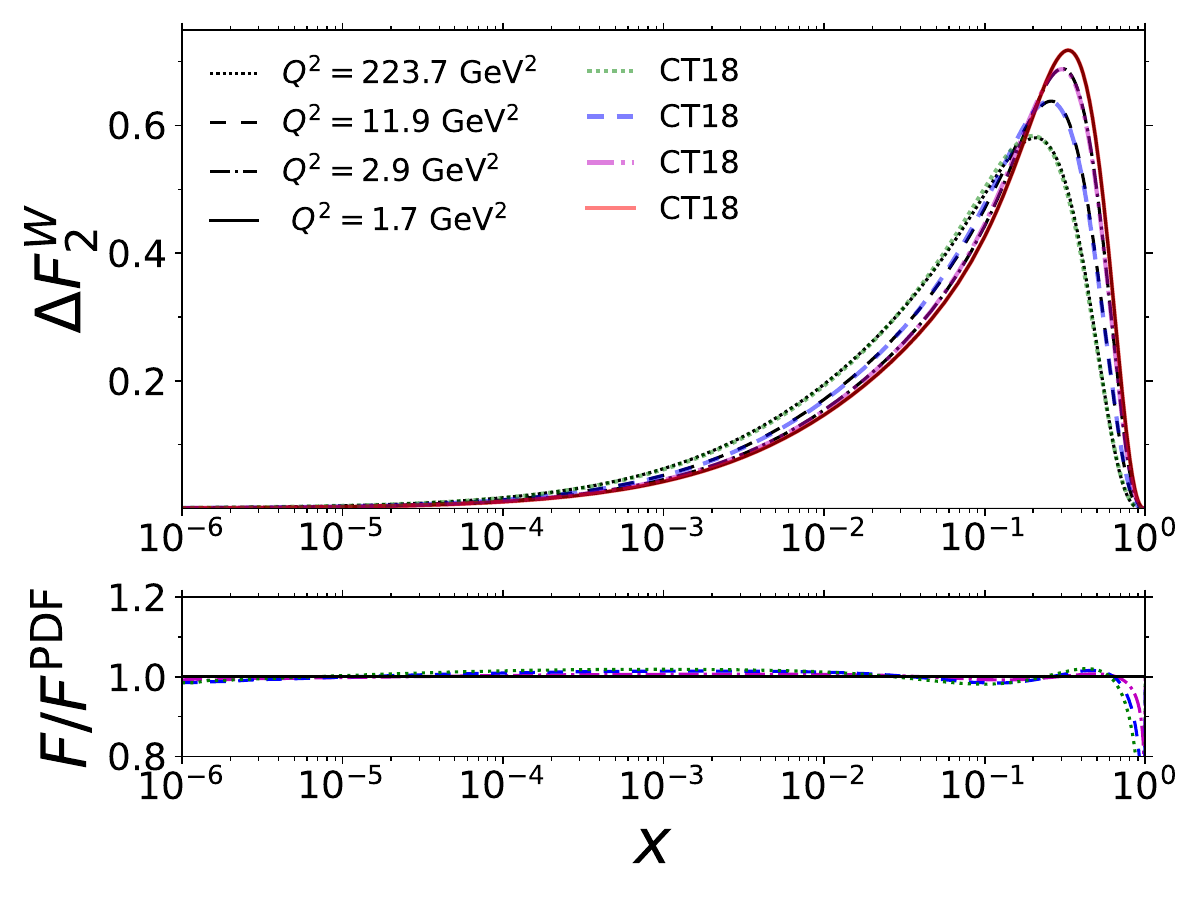}
  \includegraphics[width=.5\linewidth]{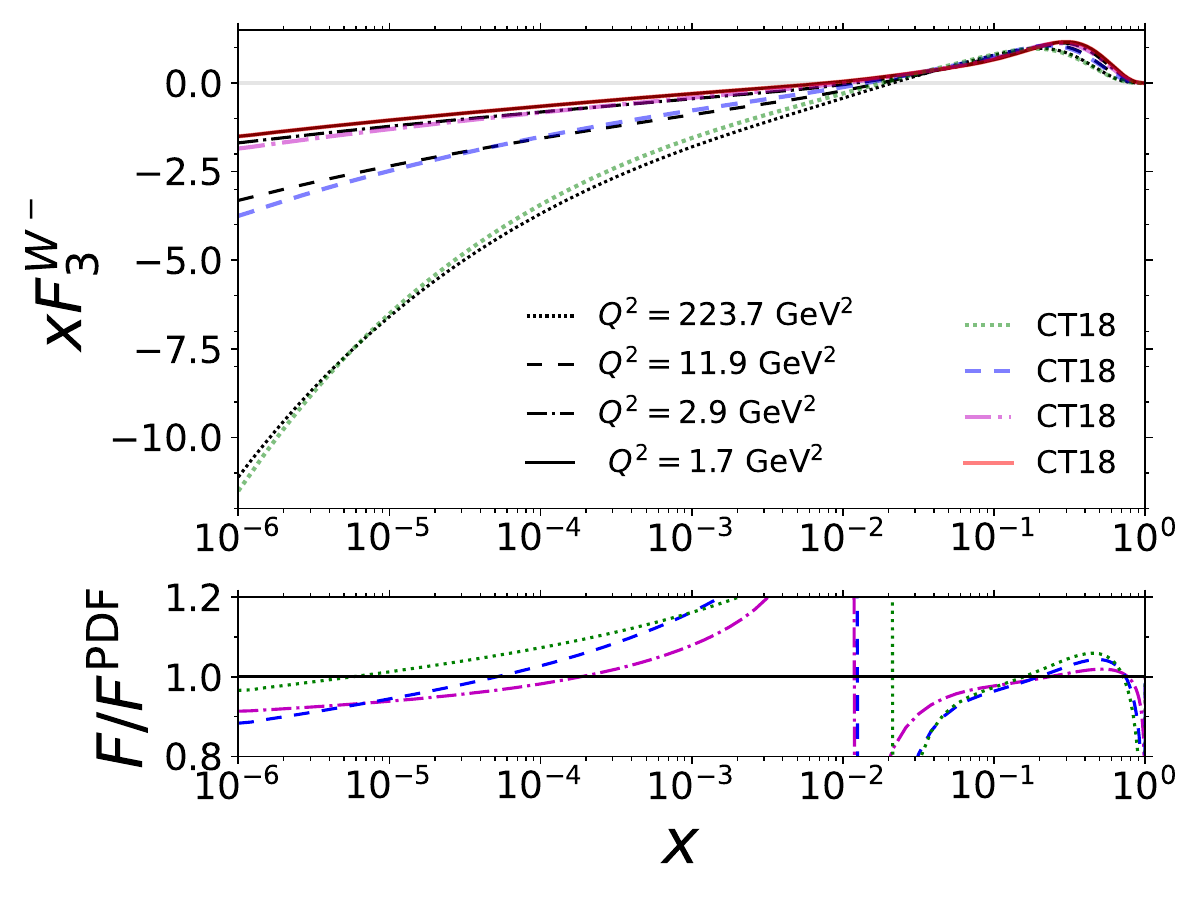}\hfill
  \includegraphics[width=.5\linewidth]{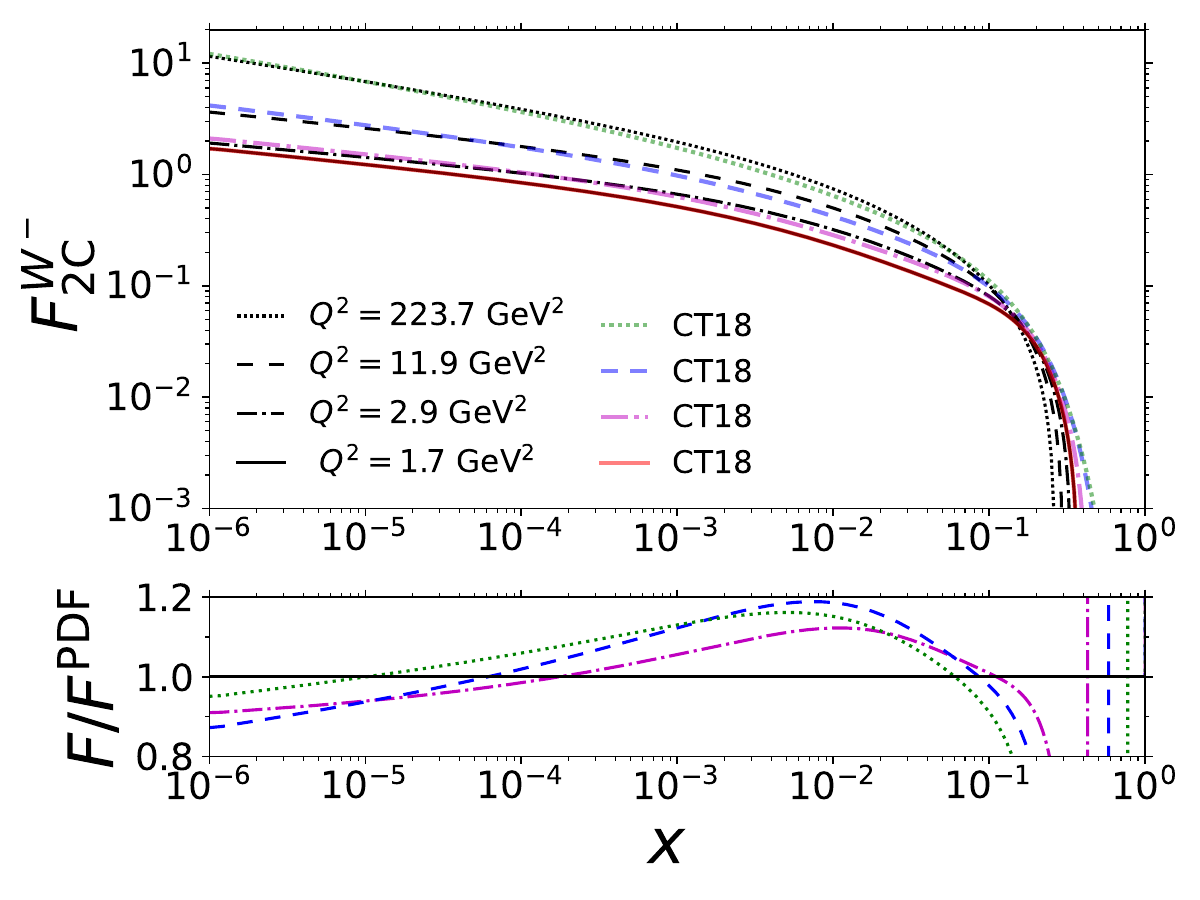}
  
  \caption{
  The $x$ dependence of $\ft$, $\fl$, $\fk$, $\ftwdelta$, $\fkw$, and $\ftcw$ using the physical-basis approach (black curves) compared with the usual PDF-based approach (colourful curves).}
\label{fig:xresultsFullBasis}
\end{figure}

\begin{figure}[ht]
  \includegraphics[width=.5\linewidth]{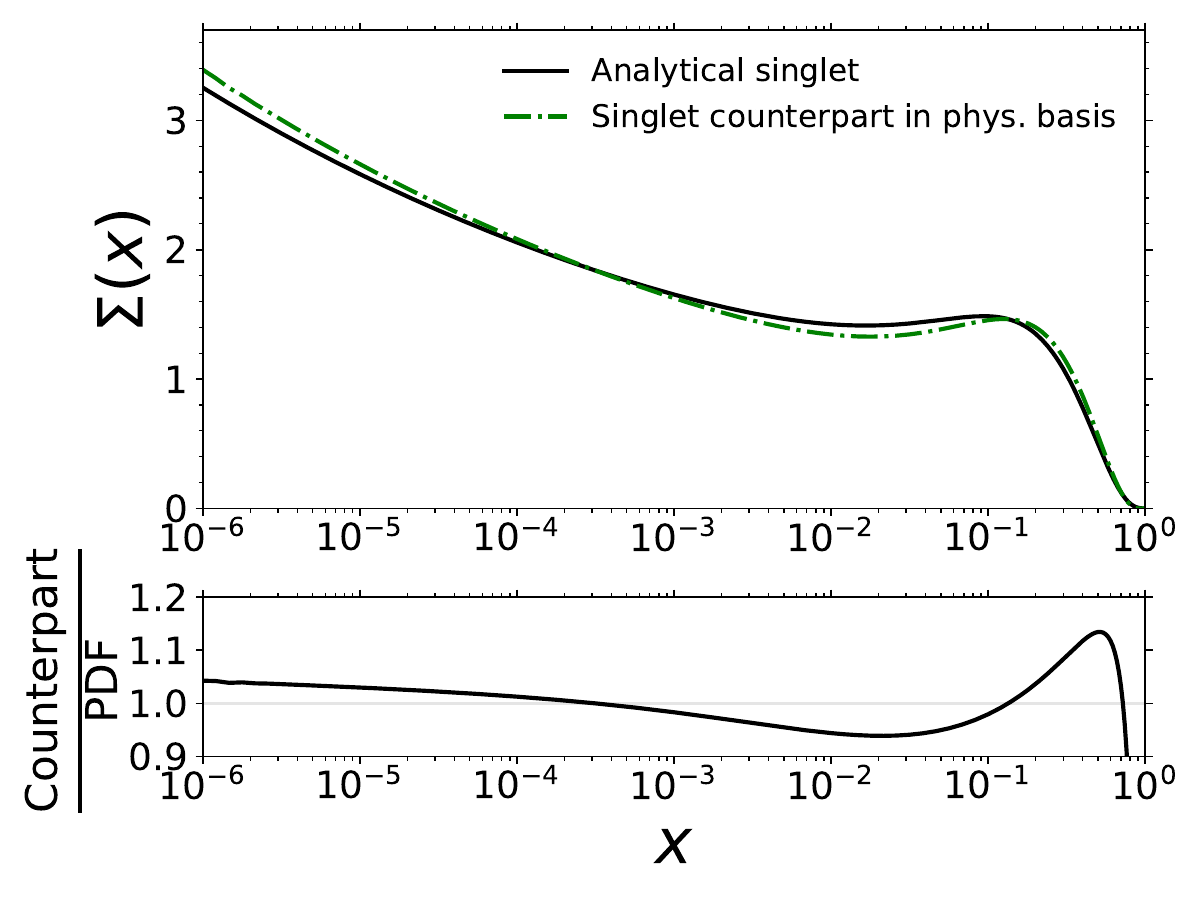}\hfill
  \includegraphics[width=.5\linewidth]{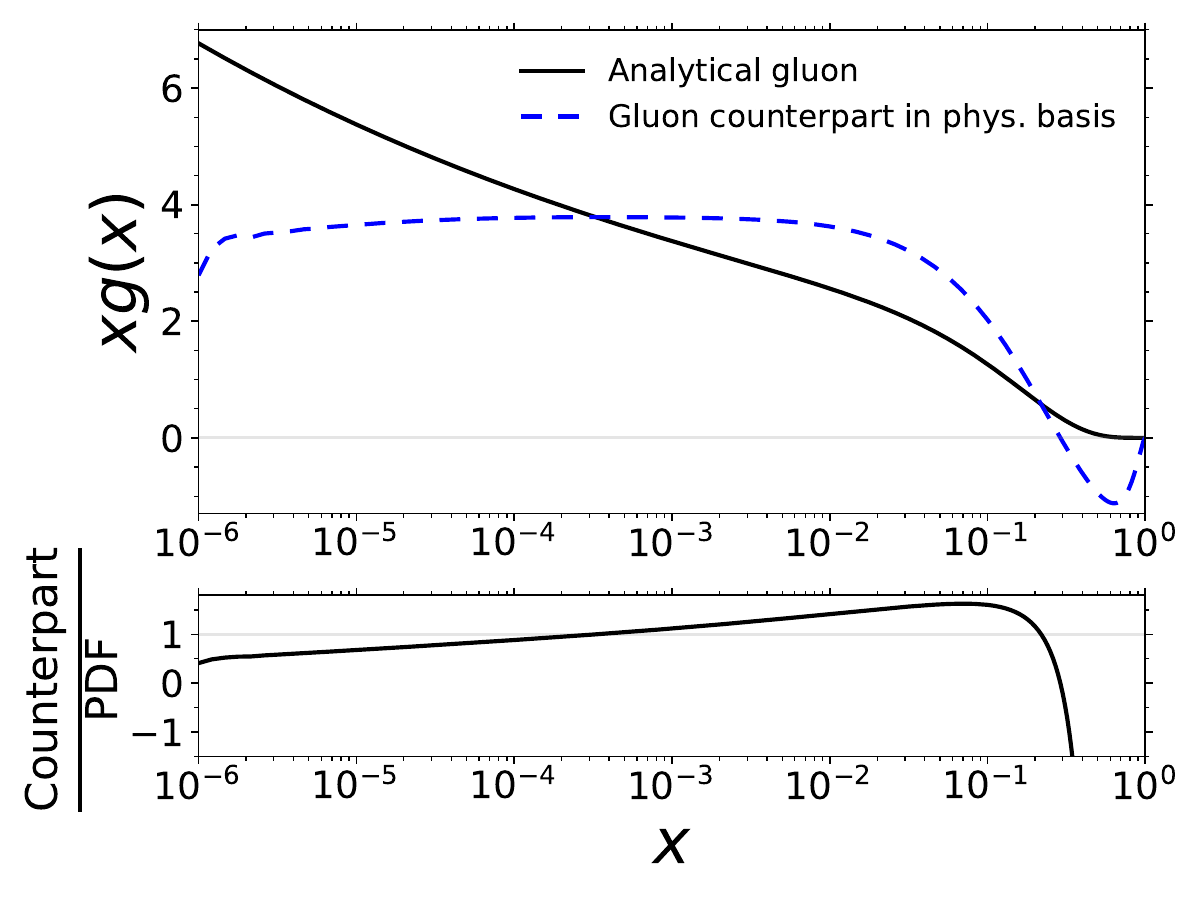}
  \caption{The NLO quark singlet and the gluon PDF counterparts in the physical basis (dashed colourful lines) where the structure functions have been computed by using analytical PDFs (solid black lines) which fulfil the momentum sum rule. Here the value for the running coupling is $\as=0.369$. }
\label{fig:gluon and singlet}
\end{figure}

\begin{figure}[ht]
  \includegraphics[width=.5\linewidth]{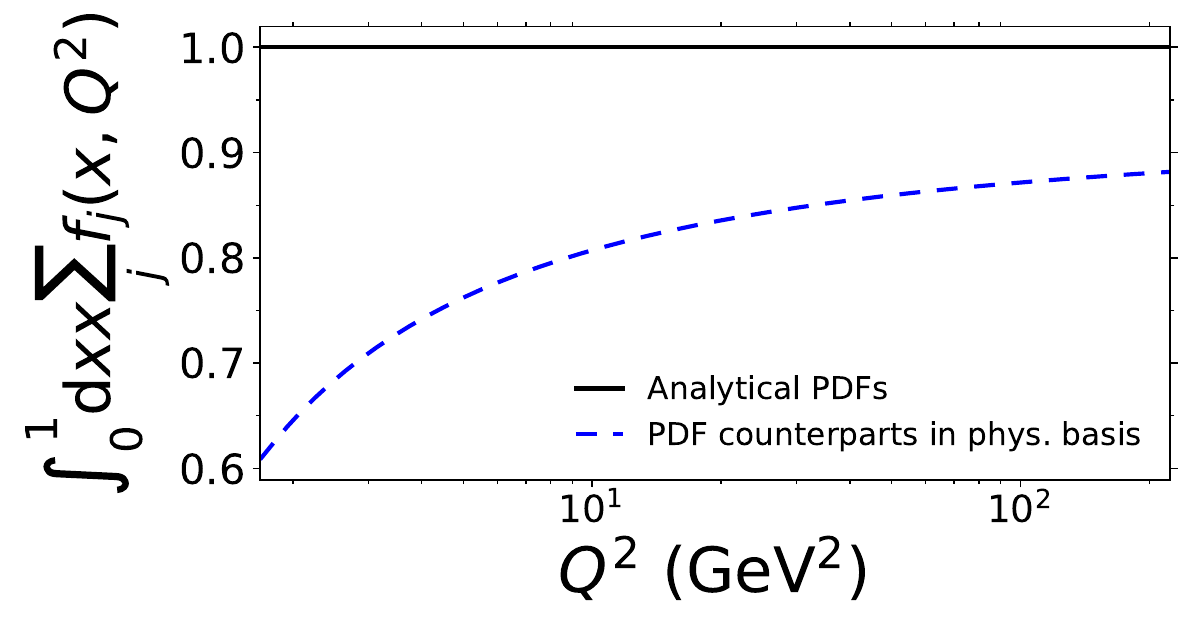}\hfill
  \includegraphics[width=.5\linewidth]{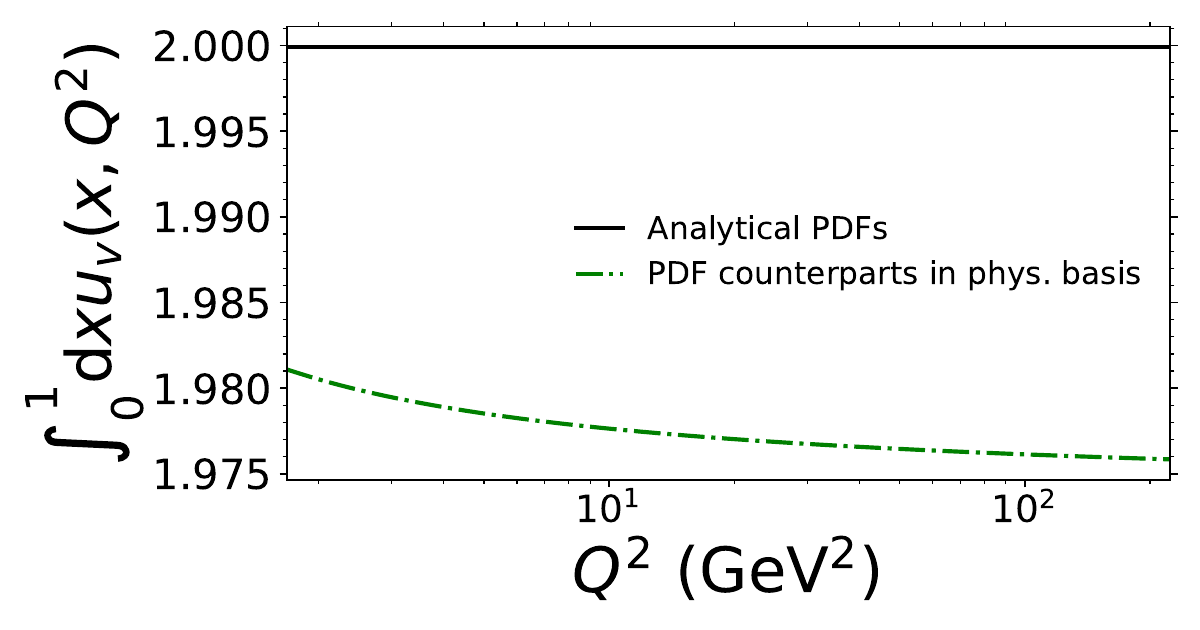}
  \includegraphics[width=.5\linewidth]{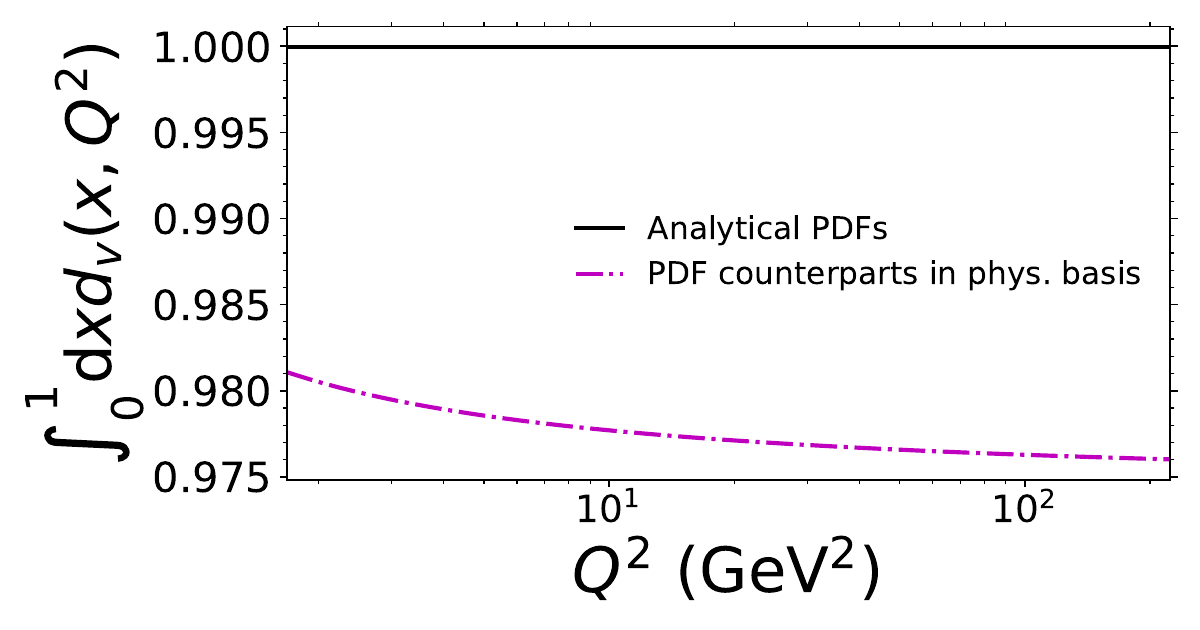}\hfill
  \caption{The momentum sum rule and the baryon number conservation rules, for the $u$ and $d$ flavours. The values computed with the physical-basis counterparts for the NLO PDFs, where the structure functions have been computed by using analytical PDFs, are shown with dashed colourful lines. And the corresponding values computed directly from the analytical PDFs are shown with solid black lines.}
\label{fig:resultsSumRule}
\end{figure}

\begin{figure}[ht]
  \includegraphics[width=.5\linewidth]{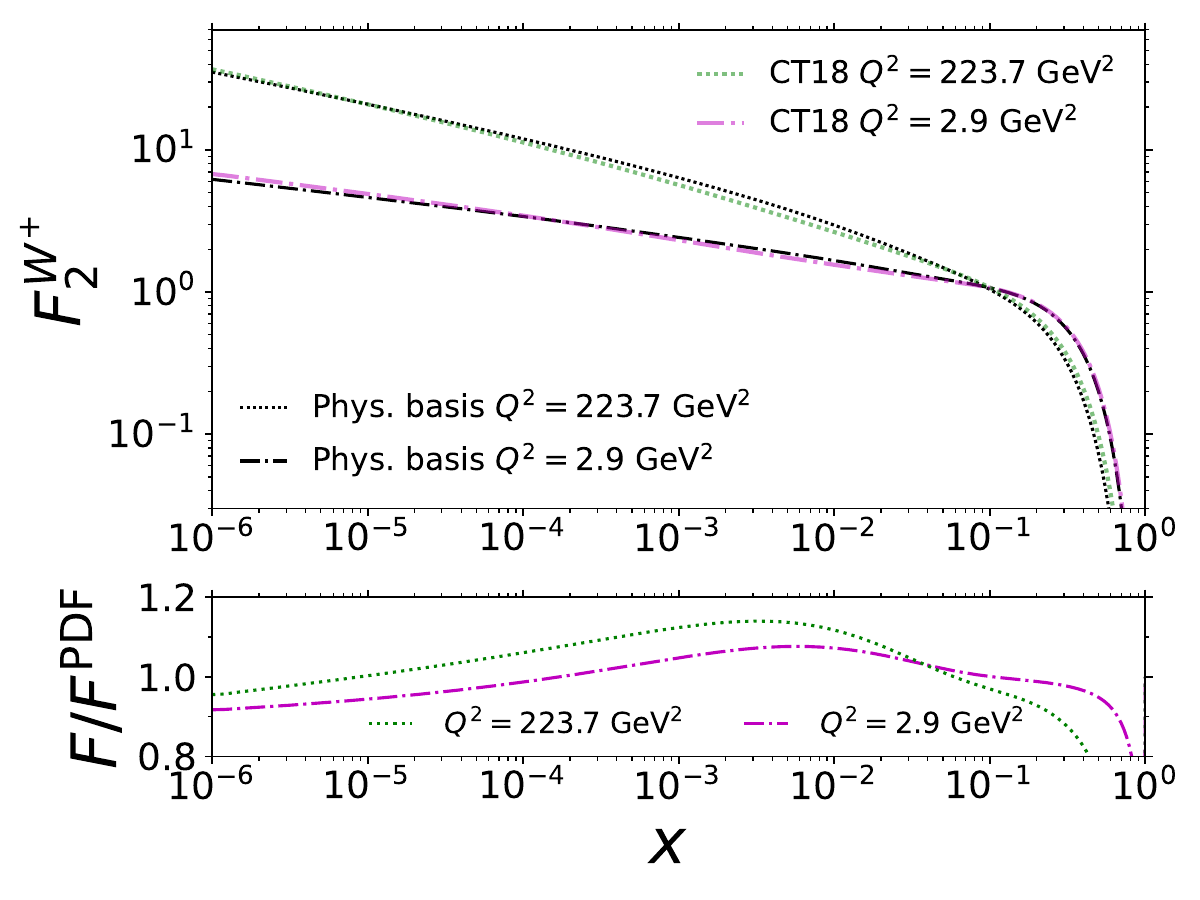}  
  \caption{The NLO structure function $F_2^{W^+}$ computed by using the PDF set "CT18NLO\_NF3" (colourful lines) and from the physical-basis expressions for the PDFs (black lines).}
\label{fig:F2Wplus}
\end{figure}

\section{Numerical results}
\label{sec:num results}
We have implemented the physical-basis DGLAP evolution numerically by using the expressions of the evolution kernels listed in Appendix~\ref{appendix Evolution kernels}. The initial conditions, the  structure functions at $Q^2=1.69 \, {\rm GeV}^2$, have been obtained by using the CTEQ PDF set ``CT18NLO\_NF3''~\cite{Hou:2019efy} and the $\msbar$ definitions in Eqs.~\eqref{eq:F2fullB}--\eqref{eq:FLfullB}. The PDF set is accessed with the LHAPDF library~\cite{Buckley_2015}. For the strong coupling $\as$ we used the values given by the CTEQ PDF set, which correspond to the NLO renormalization group equation in Eq.~\eqref{eq: renormalization group eq}. The comparisons between the structure functions computed from the PDFs and with the physical basis are shown in Fig.~\ref{fig:Q2resultsFullBasis} as functions of~$Q^2$. 
Overall, the physical basis results are not equal to, but also do not deviate dramatically from the PDF evolution ones. This is to be expected, since they are equal up to terms that are higher order in $\as$.

Additional comparisons highlighting the  behaviour as a function of $x$ are shown in Fig.~\ref{fig:xresultsFullBasis}. 
The relative differences in Fig.~\ref{fig:xresultsFullBasis} increase rapidly at values of $x$ where the structure functions approach  zero, such as at $x\to 1$, but the absolute differences remain small.
Otherwise, the structure functions evaluated with physical-basis DGLAP evolution behave similarly to the PDF-based structure functions, the differences are on the order of $10~\%$. The differences are particularly small, only a few  per cent, with the structure functions $\fk$ and $\ftwdelta$ which are linear combinations only of the valence quarks. Their physical-basis DGLAP evolution in Eqs.~\eqref{eq: kernels F3 DGLAP} and \eqref{eq: kernels deltaF2 DGLAP} is self-contained; their evolution equations do not involve other structure functions. The DGLAP evolution in the physical basis tends to generate 
smaller structure functions at $x\to 1$
than the conventional evolution. This can be seen in Fig.~\ref{fig:xresultsFullBasis} where the structure functions corresponding to the evolution in the physical basis are smaller than PDF-based values at large $x$. The differences between the PDF-based evolution and the physical-basis evolution visible here arise technically from the perturbative truncation in solving the PDF counterparts from the structure functions, which leads to deviations in the evolution that are formally of higher order in $\as$. However, we stress that given the initial conditions, the physical-basis result is unique while the results with PDFs would vary depending on the choice of the factorization scale and scheme. To demonstrate this dependence, one would need to construct PDFs with different choices of factorization scales and schemes that correspond to the same set of structure functions at the initial scale but whose predictions at higher $Q^2$ would be mutually different. The width of the ``error band'' given by such a procedure would depend on how large deviations from the default choice $\mu_f^2=Q^2$ and $\overline{\rm MS}$ one would consider.

While our main focus is on the observable physical basis structure functions, it can be instructive to also look at the ``inverted'' (i.e. calculated from the structure functions) PDFs. Since the expressions for the inverted PDFs are truncated at the second non-zero order in $\as^2$, they do not exactly match  the original PDFs. This is demonstrated in Fig.~\ref{fig:gluon and singlet} where we plot the gluon PDF and the quark singlet taken from an analytical parametrization and the corresponding NLO counterparts obtained from Eqs.~\eqref{eq:gluon PB full} and \eqref{eq:singlet PB NLO}. The value for the running coupling in this comparison is relatively large $\as=0.369$, meaning that the NLO effects in the PDF values are relevant. The analytical PDFs in this plot are same as in Ref.~\cite{Hentschinski:2013zaa}, but the $\overline{s}$ has been multiplied with an additional factor of $0.2635$ in order to enforce the momentum sum rule. In the case of the quark singlet, the differences between the original and inverted quark singlet are mostly within 10\,\%. 
The effect on the gluon PDF, on the other hand, is quite large, especially at the large~$x$ values where the physical-basis inverted  NLO gluon has negative values. The negative values at large $x$ are caused by the derivatives, in particular acting on $\fl$, in $\Gnlo$ in Eq.~\eqref{eq:gluon PB full}.
We emphasize that the physical-basis inverted NLO gluon distribution plotted in Fig.~\ref{fig:gluon and singlet} never appears in a calculation of any physical observable as such. Rather, what is plotted is a sum of the  terms $\Glo$ and $\Gnlo$ that in a cross section calculation are separately convoluted with the appropriate parts of the coefficient functions, in order to obtain a factorization-scheme independent result at a consistent order in $\as$ (see e.g. Eq.~\eqref{eq:genericxs} below). Nevertheless, the behaviour seen in Fig.~\ref{fig:gluon and singlet}
explains why the difference between physical-basis and PDF evolution for the structure functions $\fk$ and $\ftwdelta$ is smaller than for the structure functions whose evolution   contains the counterpart for the gluon PDF. Incidentally, in  Ref.~\cite{Candido:2023ujx} it was discussed that the physical-basis approach can be used to show that the values of PDFs in the $\msbar$ scheme have to remain positive. It seems that the perturbative inversion of the analytical PDFs, which can lead to negative values for the gluon, does not support this argument.

In the PDF basis, the DGLAP evolution in the $\overline{\rm MS}$ scheme exactly respects the momentum and baryon number conservation sum rules. In the physical basis, since the PDFs are obtained from the structure functions only in a perturbative expansion, these sum rules are not satisfied exactly.
We demonstrate this by plotting the momentum sum and the number of valence quarks calculated within the physical-basis approach in Fig.~\ref{fig:resultsSumRule}. In these plots, the initial condition at low $Q^2$ is taken from the analytical PDFs discussed above, and the evolution to higher $Q^2$ is carried out within the physical basis. Here, the inverted NLO gluon PDF has a particularly significant effect. The negative values of the inverted gluon at large~$x$ reduce the full momentum sum in the physical basis. As can be seen, the momentum sum neither remains constant towards higher $Q^2$. We emphasize that this is an issue that only shows up in a significant way in  PDFs which -- neither the original or physical-basis counterparts -- are not observables. The sum rule is conserved perturbatively, up to a specific order in $\as$, which is to be expected in a perturbative calculation. The effect of the momentum sum rule on physical observables only shows up at very large $x$. Physical observables directly sensitive to the baryon number density $\fk,\ftwdelta$ show negligible differences to the conventional PDF approach in the $x$ region where the bulk of the baryon number is carried. 
In a fit to experimental data in the physical-basis approach, it might still be desirable to prefer parametrizations that minimize the sum rule violation or perhaps enforce the sum rules at asymptotically large values of $Q^2$ where the $\alpha_{\rm s}$ becomes small and the sum-rule breaking NLO effects eventually disappear. 

The physical basis is applicable and universal in the same way as PDFs are; one can construct other cross sections in the physical basis only by replacing the PDFs with the physical-basis counterparts. This is demonstrated in Fig.~\ref{fig:F2Wplus} where we show a DIS structure function $F_2^{\rm W^+}$ (which is not an element in our basis) computed with the physical basis and with the PDFs from "CT18NLO\_NF3". The deviations from the PDF-based values are within $20~\%$, except at large~$x$.

\iffalse
\begin{equation}
\label{eq:leibniz rule}
    \frac{\dd}{\dd x} f\otimes g = -\frac{1}{x}f(x)g(1) + \int_x^1\frac{\dd z}{z} f(z) \frac{\dd}{\dd x}g\left(\frac{x}{z}\right) =  \frac{1}{x}\left[-f(x)g(1) +f\otimes z \frac{\dd}{\dd z}g\left(z\right)\right]
\end{equation}
\fi

\section{Conclusions and Outlook}
\label{sec:conclusions}
In this paper, we extend the physical-basis approach of $Q^2$ evolution for DIS structure functions to next-to-leading order in~$\as$. The procedure here closely follows the previous work~\cite{Lappi:2023lmi} for the LO physical basis for six structure functions. To construct the six observable basis at NLO, we however choose the structure functions $\ft$, $\fk$, $\fkw$, $\ftcw$, $\fl$, and $\ftwdelta=\ftwminus-\ftwplus$, from which the last is different to the LO basis. This change was made to acquire better numerical constraint for the valence-quark counterparts in the physical basis. 
At LO, a Mellin space procedure was used to find the inverse of the coefficient function $\cflglo$ for the gluon contribution to $\fl$. At NLO, such a full inversion does not need to be carried out, since the inversion from structure functions to PDFs can be carried out perturbatively via an iterative procedure using the LO inverse of the coefficient function.
The perturbative method allows for a straightforward extension to higher orders in~$\as$. 

The main result for this work are Eqs.~\eqref{eq: F2 DGLAP} -- \eqref{eq: FL DGLAP}. This closed set of  equations represents the DGLAP evolution of the set of physical basis structure functions. These equations are accurate to NLO order in the coupling $\as$.
Technically, these equations have been obtained by inserting the inverted PDFs into the conventional DGLAP evolution equations for the structure functions. The factorization scale and scheme dependence cancels between the NLO coefficient functions and splitting functions in the DGLAP evolution equations, leaving the the physical basis evolution factorization scale and scheme independent. 

We also solved the physical basis evolution equations numerically. The numerical results for the $Q^2$ evolution in the physical basis showed that the differences to the structure functions evolved with the PDFs are typically within a few tens of percent. The differences between physical-basis and PDF evolution in the NLO results are caused by the factorization scheme and scale dependence in the PDFs, along with the perturbative inversion of the PDF counterparts in the physical basis. These deviations are all higher-order effects in $\as$, thus they are expected to decrease when the physical basis is extended to higher orders in~$\as$. We demonstrated the universality of the physical basis by expressing the structure function $\ftwplus$ that is not part of our basis in terms of the inverted PDFs, as shown in Fig.~\ref{fig:F2Wplus}. The deviations caused by the perturbative truncation in the inverted PDFs were illustrated in Fig.~\ref{fig:gluon and singlet} by using an analytical parametrization of PDFs. 

In this paper we quantified the differences of the physical-basis results from the conventional approach for DIS structure functions. However, it is not obvious how these deviations will manifest in cross sections other than DIS observables e.g. cross sections for different processes at the LHC. 
Thus one focus in future work will be to study the application of the physical basis to LHC cross sections. When inserting the inverted PDFs into expressions for cross sections, the factorization scale and scheme dependences  cancel. This was discussed in more detail in  our previous work~\cite{Lappi:2023lmi}, where the argument was based on Ref.~\cite{Harland-Lang:2018bxd}. The cancellation however requires that one separates between the terms of the inverted PDFs corresponding to different perturbative orders. For example, a generic cross section at NLO would be calculated as truncated power series in~$\as$ 
\begin{align}\label{eq:genericxs}
    \sigma &= \sum_{ij} f_i\otimes\sigma_{ij}\otimes f_j \nonumber \\ &= \sum_{ij} f_i^{(0)}\otimes\sigma_{ij}^{(0)}\otimes f_j^{(0)} +\as \sum_{ij}\left[ f_i^{(0)}\otimes\sigma_{ij}^{(1)}\otimes f_j^{(0)} + f_i^{(1)}\otimes\sigma_{ij}^{(0)}\otimes f_j^{(0)}+f_i^{(0)}\otimes\sigma_{ij}^{(0)}\otimes f_j^{(1)}\right]+\mathcal{O}(\as^2) \,,
\end{align}
where $f_i^{(0)}$ and $f_i^{(1)}$ are the LO and NLO physical-basis counterparts for the PDFs. 
In contrast, the conventional approach with PDFs would include only an NLO PDF set. 
With this extra step of separating contributions at different orders in a consistent way, existing codes for computing hadronic cross sections could be used in a  relatively straightforward manner, in a similar spirit as in Ref.~\cite{Barontini:2024xgu}.

The ideal use case for the physical-basis approach would be to perform a global analysis of DIS data. However, before that can be made at a phenomenologically realistic level, the fixed 3-flavour scheme should be extended to a variable-flavour number scheme to enlarge the region of applicability of the framework to high $Q^2$ and preferentially heavy quark mass effects included as well. For example, when considering the charm quark as an active parton, we have to include one additional structure function in the physical basis. The structure function could be e.g. $F_{2 \rm c}$, which would be the structure function most sensitive to the charm content of the nucleon, whether intrinsic or perturbatively generated. By including heavy quark masses, we have to deal with logarithms of the form  $\log(Q^2/m_Q^2)$ which are not infrared safe. This issue is the same as in the case of heavy-quark PDFs. A separate complication is that in the presence of quark masses $Q^2$ appears not only in ratios to the factorization scale and in $\as(Q^2)$, but also in ratios to the quark mass, e.g. in the lower limits of the convolutions. Thus one has to be more careful in separating derivatives with respect to the factorization scale and with respect to $Q^2$, e.g. in Eq.~\eqref{eq: structure function DGLAP in PB general notation}.  
The additional mass-dependent terms will also complicate the process of expressing PDFs in terms of structure functions. For example, already the first non-zero coefficient function $\cflglo$ in the definition of $\fl$ will include additional mass-dependent terms~\cite{Gluck:1979aw} and the differential operator $\hat{P}$, defined in Eq.~\eqref{eq:Phat} will no longer be an exact inverse of $\cflglo$ as in Eq.~(\ref{eq:invertG}), but will need to be reworked in a suitable way.  
Thus, although the physical-basis approach is currently limited to the next-to non zero order in~$\as$ and to low $Q^2$ due to absence of heavy-flavour contributions, there are prospects for several further developments that can extend the physical basis towards becoming a full alternative to the PDF evolution approach. 

\begin{acknowledgments}
This work was supported by the Academy of Finland, the Centre of Excellence in Quark Matter (projects 346324 and 346326), projects 338263, 346567 and 359902 (H.M), projects 321840 (T.L, M.T), and project 308301 (H.P., M.T). This work was also supported under the European Union’s Horizon 2020 research and innovation programme by the European Research Council (ERC, grant agreements ERC-2023-COG-101123801 GlueSatLight and ERC-2018-ADG-835105 YoctoLHC). The content of this article does not reflect the official opinion of the European Union and responsibility for the information and views expressed therein lies entirely with the authors. 
\end{acknowledgments}

\pagebreak
\appendix
\section{Coefficient functions and DGLAP splitting functions}
\label{appendix C and P}
The LO coefficient functions in the structure function $\fl$ read as
\begin{align}
    \label{eq: CFLqLO}
    \cflqlo(x) =& 2\cf x \,, \\
    \label{eq: CFLgLO}
    \cflglo(x) =& 4\TR x(1-x) \,,
\end{align}
where $\cf = 4/3$.
The corresponding NLO coefficient functions in the $\msbar$ are
\begin{align}
    \label{eq: CFLnsNLO}
    \cflnsnlo(x) =&  \frac{1}{4}\Bigg\{ \frac{128}{9}x\log\left(1-x\right)^2 - 46.50x\log\left(1-x\right)-84.094\log(x)\log\left(1-x\right)-37.338+89.53 x +33.82x^2 \nonumber \\ &+x\log(x)(32.90+18.41\log(x))-\frac{128}{9}\log(x)-0.012\delta\left(1-x\right) \nonumber \\ &+\frac{16}{27}\nf \left[6x\log\left(1-x\right)-12x\log(x)-25x+6\right] \Bigg\} \,,\\
    \label{eq: CFLpsNLO}
    \cflpsnlo(x) =&  \frac{\nf}{4}\Bigg\{ (15.94-5.212x)\left(1-x\right)^2\log\left(1-x\right)+(0.421+1.520 x)\log(x)^2+28.09\left(1-x\right)\log(x)\nonumber\\& -\left(\frac{2.370}{x}-19.27\right)\left(1-x\right)^3 \Bigg\} \,, \\
    \label{eq: CFLgNLO}
    \cflgnlo(x) =&  \frac{1}{8}\Bigg\{ (94.74-49.20 x)(1-x)\log(1-x)^2+864.8(1-x)\log(1-x)+1161 x \log(1-x)\log(x)\nonumber \\ & +60.06 x \log(x)^2+39.66 (1-x)\log(x)-5.333\left(\frac{1}{x}-1\right)\Bigg\} \,,
\end{align}
where we use a slightly  different normalization convention than  the original source \cite{Moch:2004xu}.

The $\msbar$ NLO coefficient functions in the definitions of the structure functions $\ft$, $\fk$, $\ftwdelta$, $\fkw$, and $\ftcw$ are
\begin{align}
    \label{eq: CF2qNLO}
    \cftqnlo(x) =& \cf\left\{3 + 2x - \log(x)\frac{1+x^2}{1-x} -\frac{3}{2}\frac{1}{(1-x)_+}+ (1 + x^2)\left(\frac{\log(1-x)}{1-x} \right)_+ - \left(\frac{9}{2} +\frac{\pi^2}{3}\right) \delta(1-x)\right\} \,,\\
    \label{eq: CF2gNLO}
    \cftgnlo(x) =&  \TR\left\{(x^2 + (1 - x)^2)\log\left(\frac{1-x}{x}\right) - 1 + 8 x (1 - x)\right\}\,,\\
    \label{eq: CF3qNLO}
    \cfkqnlo(x) =&  \cftqnlo(x) -\cf(1+x)\,.
\end{align}

The LO DGLAP splitting functions read as
\begin{align}
\label{eq:Pqqlo}
\Pqqlo(x) & = \cf\left[\frac{1+x^2}{(1-x)_{+}}+\frac{3}{2}\delta(1-x) \right] \,, \\ 
\label{eq:Pqglo}
\Pqglo(x) & = \TR \left[x^2+(1-x)^2\right]  \,, \\
\label{eq:Pgglo}
\Pgglo(x) & =  2\nc \left[\frac{1-x}{x}+x(1-x)+\frac{x}{(1-x)_{+}}\right] + b_0\delta(1-x)  \,, \\
\label{eq:Pgqlo}
\Pgqlo(x) & = \cf\frac{1+(1-x)^2}{x} \,,
\end{align}
where $\nc=3$. The NLO splitting functions are taken from Ref.~\cite{Ellis:1996mzs}. The NLO splitting functions in the $\msbar$ scheme corresponding to the quark splitting are
\begin{align}
\label{eq:PqqVnlo}
\PVqqnlo(x) & =  \cf^2\Bigg\{-\left[2\log(x)\log(1-x) + \frac{3}{2}\log(x)\right](-1-x) - \left(\frac{3}{2} +\frac{7}{2}x\right)\log(x) - \frac{1}{2}(1 + x)\log(x)^2 - 5(1 - x) \Bigg\}\nonumber\\&
+\cf\nc\Bigg\{  \left[ \frac{1}{2}\log(x)^2 + \frac{11}{6}\log(x) + \frac{67}{18} - \frac{\pi^2}{6}\right](-1-x) + (1 + x)\log(x) + \frac{20}{3}(1 - x) \Bigg\}\nonumber\\& 
+\cf \tf\Bigg\{ -\left[\frac{2}{3}\log(x) + \frac{10}{9}\right](-1-x) - \frac{4}{3}(1 - x) \Bigg\} \nonumber\\& +\Bigg\{\cf^2\left[\frac{3}{8} - \frac{\pi^2}{2} + 6 \zeta(3)\right] + \cf\nc\left[\frac{17}{24} + \frac{11\pi^2}{18} - 3 \zeta(3)\right] - \cf \tf\left[\frac{1}{6} + \frac{2\pi^2}{9}\right]\Bigg\}\delta(1-x) \nonumber \\& +\Bigg\{  -\cf^2\left[2\log(x)\log(1-x) + \frac{3}{2}\log(x)\right] + \cf\nc\left[\frac{1}{2}\log(x)^2 + \frac{11}{6}\log(x) + \frac{67}{18}- \frac{\pi^2}{6}\right]\nonumber\\&
-\cf \tf \left[\frac{2}{3}\log(x)+\frac{10}{9}\right]\Bigg\}\frac{2}{(1-x)_+} \,, \\ 
\label{eq:PqqbarVnlo}
\PVqqbarnlo(x) & = \cf\left(\cf-\frac{\nc}{2}\right)\left\{ 2\left( \frac{2}{1+x} -1+x\right)S_2(x)+2(1+x)\log(x)+4(1-x)\right\} \,, \\ 
\label{eq:Pqqnlo}
\Pqqnlo(x) & = \cf^2\Bigg\{-1 + x + \left(\frac{1}{2} - \frac{3}{2}x\right) \log(x) - \frac{1}{2} (1 + x) \log(x)^2 - \left[\frac{3}{2} \log(x) + 2 \log(x) \log(1-x)\right](-1-x) \nonumber\\&+ 2 \left(\frac{2}{1+x}-1+x\right)S_2(x) \Bigg\}\nonumber\\&
  + \cf\nc\Bigg\{\frac{14}{3}(1-x)+ \left[\frac{11}{6}\log(x) + \frac{1}{2}\log(x)^2 + \frac{67}{18} - \frac{\pi^2}{6}\right] (-1-x) - \left(\frac{2}{1+x}-1+x\right)S_2(x)\Bigg\} \nonumber\\&
  + \cf \tf \Bigg\{-\frac{16}{3}+ \frac{40}{3}x + \left(10 x + \frac{16}{3}x^2 + 2\right) \log(x) -\frac{112}{9} x^2 +\frac{40}{9 x}  - 2 (1 + x) \log(x)^2\nonumber\\& - \left(\frac{10}{9} + \frac{2}{3}\log(x)\right)(-1-x) \Bigg\}\nonumber\\&+\Bigg\{\cf^2\left[\frac{3}{8} - \frac{\pi^2}{2} + 6 \zeta(3)\right] + \cf\nc\left[\frac{17}{24} + \frac{11\pi^2}{18} - 3 \zeta(3)\right] - \cf \tf\left[\frac{1}{6} + \frac{2\pi^2}{9}\right]\Bigg\}\delta(1-x) \nonumber \\& 
 +\Bigg\{ -\cf^2\left[\frac{3}{2} \log(x) + 2 \log(x) \log(1-x)\right]
  + \cf\nc\left[\frac{11}{6}\log(x) + \frac{1}{2}\log(x)^2 + \frac{67}{18} - \frac{\pi^2}{6}\right]  
  \nonumber\\& - \cf \tf \left(\frac{10}{9} + \frac{2}{3}\log(x)\right) \Bigg\} \frac{2}{(1-x)_+}\,, \\
\label{eq:Pgqnlo}
\Pgqnlo(x) & = \cf^2\Bigg\{-\frac{5}{2} -\frac{7}{2}x + \left(2 +\frac{7}{2}x \right)\log(x) - \left(1 - \frac{1}{2}x\right)\log(x)^2 - 2 x \log(1-x) \nonumber\\ & - (3 \log(1-x) + \log(1-x)^2) \frac{1+(1-x)^2}{x}\bigg\}
  + \cf\nc\Bigg\{\frac{28}{9} +\frac{65}{18}x +\frac{44}{9}x^2 - \left(12 + 5 x + \frac{8}{3} x^2\right)\log(x) \nonumber \\& + (4 + x)\log(x)^2 + 2 x \log(1-x) + S_2(x)\frac{-1-(1+x)^2}{x} \nonumber \\&+ \left(\frac{1}{2} - 2 \log(x) \log(1-x) + \frac{1}{2}\log(x)^2 + \frac{11}{3} \log(1-x) + \log(1-x)^2 - \frac{\pi^2}{6}\right)\frac{1+(1-x)^2}{x}\Bigg\}
 \nonumber\\& + \cf \tf\left\{-\frac{4}{3}x- \left(\frac{20}{9} + \frac{4}{3} \log(1-x)\right)\frac{1+(1-x)^2}{x}\right\} \,,
\end{align}where $\tf=\TR\nf=3/2$ and 
\begin{equation}
    \label{eq: S2}
    S_2(x)=-2 \rm Li_2(-x) +\frac{1}{2}\log(x)^2-2\log(x)\log(1+x)-\frac{\pi^2}{6}\,.
\end{equation}
The NLO splitting function $\Psingletnlo$ appearing in the DGLAP evolution equations \eqref{eq: F2 DGLAP}--\eqref{eq: FL DGLAP} is given by
\begin{equation}
    \label{eq: PqqS}
    \Psingletnlo=\frac{1}{2\nf}\left[ \Pqqnlo-\PVqqnlo-\PVqqbarnlo \right]\,.
\end{equation}
Frequently occurring combinations of the NLO splitting functions  $\PVqqnlo$ and $\PVqqbarnlo$ are defined as
\begin{align}
    \label{eq: Pqq+}
\Pplusnlo &\equiv \PVqqnlo + \PVqqbarnlo \,, \\ 
    \label{eq: Pqq-}
\Pminusnlo  &\equiv \PVqqnlo - \PVqqbarnlo\,.
\end{align}

The NLO splitting functions in the $\msbar$ scheme corresponding to the gluon splitting are
\begin{align}
\label{eq:Pggnlo}
\Pggnlo(x) & =  \cf \tf \Bigg\{ -16+8x+\frac{20}{3}x^2+\frac{4}{3 x}-(6+10 x)\log(x)-(2+2x)\log(x)^2\Bigg\} \nonumber \\&
+\nc \tf\Bigg\{ 2-2x+\frac{26}{9}\left(x^2+\frac{1}{x} \right)-\frac{4}{3}(1+x)\log(x)-\frac{20}{9}\left( \frac{1}{x}-2+x(1-x)\right) \Bigg\} \nonumber \\&
+\nc^2\Bigg\{ \frac{27}{2}(1-x)+\frac{67}{9}\left(x^2-\frac{1}{x}\right)-\left(\frac{25}{3}-\frac{11}{3}x+\frac{44}{3}x^2 \right)\log(x)+4(1+x)\log(x)^2 \nonumber\\ &+2\left( -\frac{1}{x}-2-x(1+x)+\frac{1}{1+x}\right)S_2(x)+\left[ \frac{67}{9}-4\log(x)\log(1-x)+\log(x)^2-\frac{\pi^2}{3}\right]\left( \frac{1}{x}-2+x(1-x)\right) \Bigg\}\nonumber \\ & + \left\{ \nc^2\left[\frac{8}{3}+3\zeta(3)\right] - \cf \tf-\frac{4}{3}\nc \tf\right\}\delta(1-x)\nonumber\\ & +\Bigg\{ -\frac{20}{9}\nc \tf+\nc^2\left[ \frac{67}{9}-4\log(x)\log(1-x)+\log(x)^2-\frac{\pi^2}{3}\right]\Bigg\}\frac{1}{(1-x)_+} \,, \\
\label{eq:Pqgnlo}
\Pqgnlo(x) & =\frac{1}{2\nf}\Bigg\{\cf \tf\Bigg[4 - 9x - (1 - 4 x) \log(x) - (1 - 2 x)\log(x)^2 + 
      4 \log(1-x) \nonumber\\&+ \left[2\log\left(\frac{1-x}{x}\right)^2 - 4 \log\left(\frac{1-x}{x}\right) -\frac{2\pi}{3} + 10\right] (x^2+(1-x)^2)\Bigg] \nonumber \\&
      + \nc \tf\Bigg[\frac{182}{9} + \frac{14}{9}x +\frac{40}{9x} + \left(\frac{136}{3}x -\frac{38}{3}\right)\log(x) -  4 \log(1-x) - (2 + 8 x)\log(x)^2 + 2 S_2(x) (x^2+(1+x)^2)\nonumber\\& + \left[-\log(x)^2 + \frac{44}{3}\log(x) - 2\log(1-x)^2 + 4 \log(1-x) + \frac{ \pi^2}{3}- \frac{218}{9}\right](x^2+(1-x)^2)\Bigg]
      \Bigg\}
     \,,
\end{align}
where we have divided out the factor $2\nf$ from the splitting function $\Pqgnlo$ definition in Ref.~\cite{Ellis:1996mzs}.

\section{PDFs in the physical basis}
\label{appendix PDFs the in physical basis}

The LO PDFs for the light quark flavours, expressed in the physical basis in terms of the structure functions $\ft$, $\fk$, $\ftwdelta$, $\fkw$, and $\ftcw$, are given by
\begin{align}
    \label{eq:dLO}
    D^{(0)}\xq & =\frac{1}{4(A_d + A_u)(\ed + \eu)} \Big\{ 2(A_d +  A_u)\wft  + (\ed + 3 \eu) \fk + (-A_u(\ed +2\eu) + A_d\eu )\wftwdelta  \nonumber \\ & -2(A_d+ A_u)\eu\fkw -2( A_d + A_u)(\es+ \eu)\wftcw\Big\}  \,, \\
    \label{eq:dbarLO}
    \overline{D}^{(0)} \xq & = \frac{1}{4(A_d + A_u)(\ed + \eu)} \Big\{ 2(A_d +A_u)\wft + (\eu-\ed)\fk + (A_u \ed + A_d \eu)\wftwdelta \nonumber \\
    &-2(A_d+ A_u)\eu\fkw - 2(A_d+A_u)(\es+\eu)\wftcw   \Big\} \,, \\
    \label{eq:uLO}    
    U^{(0)}\xq & = \frac{1}{4(A_d + A_u)(\ed + \eu)} \Big\{ 2(A_d +A_u) \ft+ (\eu-\ed) \fk \nonumber\\& +(A_u\ed + A_d\eu)\wftwdelta +  2(A_d + A_u)\ed\fkw \Big\}\,, \\   
    \label{eq:ubarLO} 
   \overline{U}^{(0)}\xq & = \frac{1}{4(A_d + A_u)(\ed + \eu)} \Big\{ 2(A_d + A_u)\wft  -(3\ed +\eu)\fk \nonumber\\& + (-A_d(2\ed+\eu) + A_u\ed)\wftwdelta +  2(A_d+A_u)\ed\fkw \Big\} \,, \\
     \overline{S}^{(0)}\xq & = S^{(0)}\xq= \frac{1}{2}\wftcw \,,
    \label{eq:sLO}
    \end{align}    
where we have defined $A_q \equiv L^2_q-R_q^2$ in order to simplify the notation. The NLO corrections are 
\begin{align}
    \label{eq:dNLO}
    D^{(1)}\xq & =
    -\frac{1}{4(A_d + A_u)(\ed + \eu)}\Big\{ 2(A_d +A_u )\cftqnlo\otimes\wft +(\ed + 3 \eu)\cfkqnlo\otimes\fk \nonumber \\ & +(-A_u(\ed +2\eu) + A_d\eu )\cftqnlo\otimes\wftwdelta  -2(A_d+ A_u)\eu\cfkqnlo\otimes\fkw  \nonumber \\ & -2 (A_d+ Au)(\es+\ed)\cftqnlo\otimes\wftcw  -  4 (A_d+A_u) (\es+ \eu -\nf\seqav)\cftgnlo\otimes G^{(0)}  \Big\} \,, \\
    \label{eq:dbarNLO}
    \overline{D}^{(1)} \xq & = -\frac{1}{4(A_d + A_u)(\ed + \eu)}\Big\{ 2(A_d +A_u)\cftqnlo\otimes\wft + (\eu-\ed)\cfkqnlo\otimes\fk\nonumber \\
    &+ (A_u \ed + A_d \eu)\cftqnlo\otimes\wftwdelta  -2(A_d+ A_u)\eu\cfkqnlo\otimes\fkw \nonumber \\
    &- 2(A_d+A_u)(\es+\eu)\cftqnlo\otimes\wftcw -4( A_d+ A_u)(\es+\eu - \nf\seqav)\cftgnlo\otimes G^{(0)}  \Big\}\,, \\
    \label{eq:uNLO}    
    U^{(1)}\xq & =- \frac{1}{4(A_d + A_u)(\ed + \eu)}\Big\{2(A_d +A_u) \cftqnlo\otimes \ft+ (\eu-\ed) \cfkqnlo\otimes\fk  \nonumber \\ &+(A_u\ed + A_d\eu)\cftqnlo\otimes\wftwdelta  +  2(A_d + A_u)\ed\cfkqnlo\otimes\fkw +4(A_d +A_u)\nf\seqav\cftgnlo\otimes G^{(0)} \Big\}  \,, \\   
    \label{eq:ubarNLO} 
   \overline{U}^{(1)}\xq & = - \frac{1}{4(A_d + A_u)(\ed + \eu)}\Big\{  2(A_d + A_u)\cftqnlo\otimes\wft  -(3\ed +\eu)\cfkqnlo\otimes\fk \nonumber \\ &+ (-A_d(2\ed+\eu) + A_u\ed)\cftqnlo\otimes\wftwdelta +  2(A_d+A_u)\ed\cfkqnlo\otimes\fkw \nonumber \\ &+4(A_d+ A_u)\nf\seqav\cftgnlo\otimes G^{(0)} \Big\} \,, \\
     \overline{S}^{(1)}\xq & = S^{(1)}\xq= -\left[  \frac{1}{2}\cftqnlo\otimes\wftcw+\cftgnlo\otimes G^{(0)} \right] \,.
    \label{eq:sNLO}
    \end{align}    
Combining these, the LO and NLO components for the quark singlet in the physical basis read
\begin{align}
    \label{eq:pure singlet LO phys basis}
    \PSlo\xq &=    \frac{1}{2(A_d + A_u)(\ed + \eu)}\Big\{ 4(A_d +A_u)\wft +  2 (\eu-\ed)\fk \nonumber \\&+(A_u-A_d)(\ed - \eu) \wftwdelta + 2(A_d + A_u)(\ed - \eu)\fkw \Big\} \,, \\
    \label{eq:pure singlet NLO phys basis}
    \PSnlo\xq &= -\frac{1}{2(A_d + A_u)(\ed + \eu)}\Big\{ 4(A_d +A_u)\cftqnlo\otimes\wft +  2 (\eu-\ed)\cfkqnlo\otimes\fk \nonumber \\&  +(A_u-A_d)(\ed - \eu) \cftqnlo\otimes\wftwdelta   + 2(A_d + A_u)(\ed - \eu)\cfkqnlo\otimes\fkw \nonumber \\& 
 +8(A_d+A_u)\nf\seqav\cftgnlo\otimes G^{(0)}  \Big\} \,.
    \end{align}
and the quark singlet weighted with the electric charge is 
\begin{align}
    \label{eq:LO singlet PB NLO}
    \singletlo\xq &= \wft \,, \\
    \label{eq:NLO singlet PB NLO}
    \singletnlo \xq &=- \left\{ \cftqnlo\otimes\wft +2\nf\seqav\cftgnlo\otimes \Glo \right\} \,.
\end{align}
The LO and NLO components of the physical-basis counterpart for the gluon PDF read as
\begin{align}
\label{eq:LO gluon PB}
    \Glo\xq = &\frac{1}{8\TR\nf\seqav}\left\{2\cf \left(\wftp-2\wft \right)+\wflpp-2\wflp+2\wfl\right\} \\
\label{eq:NLO gluon PB}
    \Gnlo\xq = & - \frac{1}{8\TR\nf\seqav}\hat{P}(x) \Bigg\{ \frac{\seqav}{2(A_d + A_u)(\ed + \eu)}\cflpsnlo\otimes\Big[ 4(A_d +A_u)\wft +  2 (\eu-\ed)\fk \nonumber \\&+(A_u-A_d)(\ed - \eu) \wftwdelta + 2(A_d + A_u)(\ed - \eu)\fkw \Big] +\cflnsnlo\otimes\wft \nonumber   \\&+ \frac{1}{4\TR}\cflgnlo\otimes\left[2\cf \left(\wftp-2\wft \right)+\wflpp-2\wflp+2\wfl \right] \nonumber \\&  - \cflqlo \otimes\left[ \cftqnlo\otimes\wft +\frac{1}{4\TR}\cftgnlo\otimes \left[2\cf \left(\wftp-2\wft \right)+\wflpp-2\wflp+2\wfl \right] \right] \Bigg\} \,.
\end{align}

\section{Evolution kernels}
\label{appendix Evolution kernels}
When partially integrating the convolutions $\Pqglo\otimes G$, $\cflglo\otimes G$, and $\cflqlo\otimes\Pqglo\otimes G$ in the DGLAP evolution equations~\eqref{eq: F2 DGLAP}--\eqref{eq: FL DGLAP} one has to consistently deal with the LO and NLO components. This ensures the cancellation of the non-trivial boundary terms $\wflp(1)$ and $\wftp(1)$, which are related to each other by 
\begin{align}
    \label{eq: FL derivative at 1}
    \wflp(1)=&-0.003\As\wftp(1)+\mathcal{O}(\as^2)\,,
\end{align}
where the factor $-0.003$ is multiplying the delta function in the coefficient function $\cflnsnlo$.
The evolution kernels listed in this section are obtained by using this relation, which is derived from the definition of $\fl$ in Eq.~\eqref{eq:FLfullB}.  

The non-zero LO evolution kernels appearing in Eqs.~\eqref{eq: kernels F2 DGLAP}--\eqref{eq: kernels FL DGLAP} are
\begin{align}
    \label{eq: PF2F2LO}
P^{(0)}_{\ft \wft}(z) &= \Pqqlo(z)+\frac{\cf}{2}\left(2(z-1)-\delta(1-z) \right)\,, \\
    \label{eq: PF2FLLO}
P^{(0)}_{\ft \wfl}(z) &=\frac{1}{4}\left(\delta(1-z)+2\int_z^1\frac{\dd \xi}{\xi}\delta(1-\xi) \right) \,, \\
    \label{eq: PF2FLpLO}
P^{(0)}_{\ft \wflp}(z) &=-\frac{1}{4}\,, \\
    \label{eq: PF3F3LO}
P^{(0)}_{\fk \fk}(z) &= P^{(0)}_{\ftwdelta \wftwdelta} = P^{(0)}_{\fkw \fkw}(z) = P^{(0)}_{\ftcw \wftcw}(z) = \Pqqlo(z)\,, \\ 
\label{eq: PF3WF2LO}
P^{(0)}_{\fkw \wft}(z) &= -P^{(0)}_{\ftcw \wft}(z)=-\frac{\cf}{2\nf\seqav}\left(2(z-1)-\delta(1-z) \right)\,, \\
 \label{eq: PF3WFLLO}
P^{(0)}_{\fkw \wfl}(z) &=-P^{(0)}_{\ftcw \wfl}(z)=-\frac{1}{\nf\seqav}P^{(0)}_{\ft \wfl}(z) \,, \\
  \label{eq: PF3WFLpLO}
P^{(0)}_{\fkw \wflp}(z) &=-P^{(0)}_{\ftcw \wflp}(z)=-\frac{1}{\nf\seqav}P^{(0)}_{\ft \wflp}(z) \,, \\
\label{eq: PFLF2LO}
P^{(0)}_{\fl \wft}(z) &=  \cf \Bigg[4\nc-2\nc\frac{1}{z}+2\left(\cf+\nc-b_0\right)z-4\nc z^2 \nonumber \\ 
&+4\left(2\nc -\cf\right)z\log\left(z\right)+4\left(\cf-\nc\right)z\log\left(1-z\right) \Bigg] + 4(A_d +A_u)\rho(z)\,,\\
\label{eq:PFLF3LO}
P^{(0)}_{\fl \fk}(z) & =  2 (\eu-\ed)\rho(z)\,,\\
\label{eq:PFLF2deltaWLO}
P^{(0)}_{\fl \wftwdelta}(z) & =(A_u-A_d)(\ed - \eu)\rho(z)\,,\\
\label{eq:PFLF3WLO}
P^{(0)}_{\fl \fkw}(z) & = 2(A_d + A_u)(\ed - \eu)\rho(z)\,,\\
\label{eq:PFLFL}
P^{(0)}_{\fl \wfl}(z) & = \frac{\cf}{2}\left(\delta(1-z) +2(1-z) \right)+\Pgglo  \,,
\end{align}
where we have defined 
\begin{equation}
    \label{eq:rho}
    \rho(z) \equiv \frac{\nf\seqav}{(A_d + A_u)(\ed + \eu)}\cflglo\otimes\Pgqlo = \frac{4\TR\cf \nf\seqav}{(A_d + A_u)(\ed + \eu)}\left[  -1+\frac{1}{3z}+\frac{2}{3}z^2-z\log\left(z\right)\right]\,.
\end{equation}
Note that in the LO results in previous work \cite{Lappi:2023lmi} we had a non-zero evolution kernel $P_{\fl \wflp}^{(0)}$ which is removed here by partial integration.

The NLO evolution kernels for the structure functions $\fk$ and $\ftwdelta$ in Eqs.~\eqref{eq: kernels F3 DGLAP} and \eqref{eq: kernels deltaF2 DGLAP} are simply 
\begin{align}
    \label{eq: PF2WdeltaF2WdeltaNLO}
P^{(1)}_{\ftwdelta \wftwdelta}(z) &= \Pminusnlo-b_0\cftqnlo \,, \\
    \label{eq: PF3FwNLO}
P^{(1)}_{\fk \fk}(z) &= \Pminusnlo-b_0\cfkqnlo \,.
\end{align}
When writing out the NLO evolution kernels for the rest of the structure functions one can avoid repetition of lengthy expressions by introducing a notation 
\begin{equation}
   C\otimes \Gnlo= \sum_i (C\otimes \Gnlo)_{F_i}\otimes F_i \,,    
\end{equation}
where $C(z)$ is an analytical function in $z \in [0,1]$ and $F_i \in \{\wft, \fk, \wftwdelta, \fkw, \wftcw, \wfl, \wftp, \wflp, \wflpp \}$. Here the expression $(C\otimes \Gnlo)_{F_i}$ refers to the term which is convoluted with $F_i$ after partially integrating the differential operator $\hat{P}$ from the convolution $C\otimes \Gnlo$. The terms proportional to $\wftp(1)$ are eliminated with Eq.~\eqref{eq: FL derivative at 1}.  

The convolution $\Pqglo\otimes\Gnlo$, appearing in Eqs.~\eqref{eq: F2 DGLAP}, \eqref{eq: F3W DGLAP}, and \eqref{eq: F2CW DGLAP}, can be expressed with 
\begin{align}
    (\Pqglo\otimes\Gnlo)_{\wft}=&-\frac{1}{8\nf\seqav}\Bigg\{ \frac{2\seqav}{(\ed+\eu)}\left[ \cflpsnlo-z\frac{\dd}{\dd z}\cflpsnlo(z)+2\int_z^1\frac{\dd \xi}{\xi}\cflpsnlo(\xi)\right] \nonumber \\ &+\cflnsnlo+2\int_z^1\frac{\dd \xi}{\xi}\cflnsnlo(\xi) -\frac{\cf}{\TR}\left[ \cflgnlo-z\frac{\dd}{\dd z}\cflgnlo(z) +2\int_z^1\frac{\dd \xi}{\xi}\cflgnlo(\xi)\right]\nonumber\\
& -2\cf\left[\cftqnlo+2(1-z)\otimes\cftqnlo \right]+\frac{2\cf^2}{\TR}\left[\cftgnlo+2(1-z)\otimes\cftgnlo \right] \Bigg\}\,, \\
    (\Pqglo\otimes\Gnlo)_{\wftp}=&-\frac{1}{8\nf\seqav}\Bigg\{ - \cflnsnlo +\frac{\cf}{2\TR}\left[ \cflgnlo-z\frac{\dd}{\dd z}\cflgnlo(z) +2\int_z^1\frac{\dd \xi}{\xi}\cflgnlo(\xi)\right] \nonumber\\&-\frac{\cf^2}{\TR}\left[\cftgnlo+2(1-z)\otimes\cftgnlo \right] \Bigg\}\,, \\
    (\Pqglo\otimes\Gnlo)_{\wfl}=&-\frac{1}{8\nf\seqav}\Bigg\{ \frac{1}{2\TR}\left[ \cflgnlo-z\frac{\dd}{\dd z}\cflgnlo(z) +2\int_z^1\frac{\dd \xi}{\xi}\cflgnlo(\xi)\right]\nonumber\\&-\frac{\cf}{\TR}\left[\cftgnlo+2(1-z)\otimes\cftgnlo \right] \Bigg\}\,, \\
    (\Pqglo\otimes\Gnlo)_{\wflp}=&-(\Pqglo\otimes\Gnlo)_{\wfl}\,, \\
    (\Pqglo\otimes\Gnlo)_{\wflpp}=&\frac{1}{2}(\Pqglo\otimes\Gnlo)_{\wfl}\,, \\(\Pqglo\otimes\Gnlo)_{\fk}=& -\frac{1}{8\nf}\frac{\eu-\ed}{(A_d+A_u)(\ed+\eu)}\left[ \cflpsnlo-z\frac{\dd}{\dd z}\cflpsnlo(z)+2\int_z^1\frac{\dd \xi}{\xi}\cflpsnlo(\xi)\right] \,,\\
    (\Pqglo\otimes\Gnlo)_{\wftwdelta}=& -\frac{1}{2}(A_u-A_d)(\Pqglo\otimes\Gnlo)_{\fk}\,,\\
    (\Pqglo\otimes\Gnlo)_{\fkw} =& -(A_d+A_u)(\Pqglo\otimes\Gnlo)_{\fk} \,.
\end{align}
Here the integrals and derivatives of the NLO coefficient functions $\cflpsnlo$, $\cflnsnlo$, and $\cflgnlo$ can be calculated analytically.

Now the NLO evolution kernels for the structure function $\ft$ read
\begin{align}
    \label{eq: PF2F2NLO}
P^{(1)}_{\ft \wft}(z) &= \Pplusnlo+\frac{4\nf\seqav}{\ed+\eu}\Psingletnlo-\frac{\cf}{\TR}\Pqgnlo+\frac{\cf}{\TR}\Pqqlo\otimes\cftgnlo+2\nf\seqav (\Pqglo\otimes\Gnlo)_{\wft}\nonumber\\&+\frac{4\nf\seqav}{\ed+\eu}\Pgqlo\otimes\cftgnlo-\frac{\cf}{\TR}\Pgglo\otimes\cftgnlo-\frac{\cf}{2}\left[\cftqnlo+2(1-z)\otimes\cftqnlo \right]-b_0\left[\cftqnlo-\frac{\cf}{\TR}\cftgnlo\right]
\,, \\
    \label{eq: PF2F2pNLO}
P^{(1)}_{\ft \wftp}(z) &=\frac{\cf}{2\TR}\left[\Pqgnlo-\Pqqlo\otimes\cftgnlo+\Pgglo\otimes\cftgnlo-b_0\cftgnlo \right] +2\nf\seqav(\Pqglo\otimes\Gnlo)_{\wftp} \,, \\
    \label{eq: PF2FLNLO}
P^{(1)}_{\ft \wfl}(z) &=\frac{1}{2\TR}\left[ \Pqgnlo-\Pqqlo\otimes\cftgnlo+\Pgglo\otimes\cftgnlo-b_0\cftgnlo \right]\nonumber \\& +\frac{1}{4}\left[\cftqnlo+2\int_z^1\frac{\dd\xi}{\xi}\cftqnlo(\xi)\right] +2\nf\seqav(\Pqglo\otimes\Gnlo)_{\wfl} \,, \\
 \label{eq: PF2FLpNLO}
P^{(1)}_{\ft \wflp}(z) &=-\frac{1}{2\TR}\left[ \Pqgnlo-\Pqqlo\otimes\cftgnlo+\Pgglo\otimes\cftgnlo-b_0\cftgnlo \right]-\frac{1}{4}\cftqnlo +2\nf\seqav(\Pqglo\otimes\Gnlo)_{\wflp} \,,\\
 \label{eq: PF2FLppNLO}
P^{(1)}_{\ft \wflpp}(z) &=\frac{1}{4\TR}\left[ \Pqgnlo-\Pqqlo\otimes\cftgnlo+\Pgglo\otimes\cftgnlo-b_0\cftgnlo \right] +2\nf\seqav(\Pqglo\otimes\Gnlo)_{\wflpp} \,, \\
   \label{eq: PF2F3NLO}
P^{(1)}_{\ft \fk}(z) &=2\nf\seqav\left[ \frac{\eu-\ed}{(A_d+A_u)(\ed+\eu)}\Psingletnlo +(\Pqglo\otimes\Gnlo)_{\fk} \right] \,, \\
    \label{eq: PF2F2WdeltaNLO}
P^{(1)}_{\ft \wftwdelta}(z) &=2\nf\seqav\left[\frac{(A_u-A_d)(\ed-\eu)}{2(A_d+A_u)(\ed+\eu)}\Psingletnlo+ (\Pqglo\otimes\Gnlo)_{\wftwdelta} \right] \,, \\
    \label{eq: PF2F3wNLO}
P^{(1)}_{\ft \fkw}(z) &=2\nf\seqav\left[\frac{\ed-\eu}{\ed+\eu}\Psingletnlo+(\Pqglo\otimes\Gnlo)_{\fkw}\right] \,.
\end{align}
For the structure function $\fkw$ the NLO evolution kernels can be written as
\begin{align}
    \label{eq: P3W2F2NLO}
P^{(1)}_{\fkw \wft}(z) &= -\frac{4}{\ed+\eu}\Psingletnlo+\frac{\cf}{\TR\nf\seqav}\Pqgnlo+\frac{\cf}{2\nf\seqav}\left[\cfkqnlo+2(1-z)\otimes\cfkqnlo \right]-2(\Pqglo\otimes\Gnlo)_{\wft}
\,, \\
    \label{eq: PF3wF2pNLO}
P^{(1)}_{\fkw \wftp}(z) &=-\frac{\cf}{2\TR\nf\seqav}\Pqgnlo -2(\Pqglo\otimes\Gnlo)_{\wftp} \,, \\
    \label{eq: PF3wFLNLO}
P^{(1)}_{\fkw \wfl}(z) &=-\frac{1}{2\TR\nf\seqav}\Pqgnlo-\frac{1}{4\nf\seqav}\left[\cfkqnlo+2\int_z^1\frac{\dd\xi}{\xi}\cfkqnlo(\xi)\right] -2(\Pqglo\otimes\Gnlo)_{\wfl} \,, \\
 \label{eq: PF3wFLpNLO}
P^{(1)}_{\fkw \wflp}(z) &=\frac{1}{2\TR\nf\seqav}\Pqgnlo+\frac{1}{4\nf\seqav}\cfkqnlo -2(\Pqglo\otimes\Gnlo)_{\wflp} \,,\\
 \label{eq: PF3wFLppNLO}
P^{(1)}_{\fkw \wflpp}(z) &=-\frac{1}{4\TR\nf\seqav}\Pqgnlo-2(\Pqglo\otimes\Gnlo)_{\wflpp} \,, \\
   \label{eq: PF3wF3NLO}
P^{(1)}_{\fkw \fk}(z) &=-\frac{2}{A_d+A_u}\PVqqbarnlo -\frac{2(\eu-\ed)}{(A_d+A_u)(\ed+\eu)}\Psingletnlo-2(\Pqglo\otimes\Gnlo)_{\fk} \,, \\
    \label{eq: PF3wF2WdeltaNLO}
P^{(1)}_{\fkw \wftwdelta}(z) &=-\frac{A_d-A_u}{A_d+A_u}\PVqqbarnlo -\frac{(A_u-A_d)(\ed-\eu)}{(A_d+A_u)(\ed+\eu)}\Psingletnlo -2(\Pqglo\otimes\Gnlo)_{\wftwdelta} \,, \\
\label{eq: PF3wF3wNLO}
P^{(1)}_{\fkw \fkw}(z) &=\Pplusnlo-\frac{2(\ed-\eu)}{\ed+\eu}\Psingletnlo-b_0\cfkqnlo-2 (\Pqglo\otimes\Gnlo)_{\fkw} \,.
\end{align}
The NLO evolution kernels for the structure function $\ftcw$ are
\begin{align}
    \label{eq: PF2cwF2NLO}
P^{(1)}_{\ftcw \wft}(z) &= \frac{4}{\ed+\eu}\left[\Psingletnlo+\Pgqlo\otimes\cftgnlo\right]-\frac{\cf}{\TR\nf\seqav}\left[\Pqgnlo-\Pqqlo\otimes\cftgnlo+\Pgglo\otimes\cftgnlo-b_0\cftgnlo \right]\nonumber\\&-\frac{\cf}{2\nf\seqav}\left[\cftqnlo+2(1-z)\otimes\cftqnlo \right]+2(\Pqglo\otimes\Gnlo)_{\wft}
\,, \\
    \label{eq: PF2cwF2pNLO}
P^{(1)}_{\ftcw \wftp}(z) &=\frac{\cf}{2\TR\nf\seqav}\left[ \Pqgnlo-\Pqqlo\otimes\cftgnlo+\Pgglo\otimes\cftgnlo-b_0\cftgnlo\right]+2(\Pqglo\otimes\Gnlo)_{\wftp} \,, \\
    \label{eq: PF2cwFLNLO}
P^{(1)}_{\ftcw \wfl}(z) &=\frac{1}{2\TR\nf\seqav}\left[ \Pqgnlo-\Pqqlo\otimes\cftgnlo+\Pgglo\otimes\cftgnlo-b_0\cftgnlo\right]\nonumber\\&
+\frac{1}{4\nf\seqav}\left[\cftqnlo+2\int_z^1\frac{\dd\xi}{\xi}\cftqnlo(\xi) \right]+2(\Pqglo\otimes\Gnlo)_{\wfl} \,, \\
 \label{eq: PF2cwFLpNLO}
P^{(1)}_{\ftcw \wflp}(z) &=-\frac{1}{2\TR\nf\seqav}\left[ \Pqgnlo-\Pqqlo\otimes\cftgnlo+\Pgglo\otimes\cftgnlo-b_0\cftgnlo\right]-\frac{1}{4\nf\seqav}\cftqnlo +2(\Pqglo\otimes\Gnlo)_{\wflp} \,,\\
 \label{eq: PF2cwFLppNLO}
P^{(1)}_{\ftcw \wflpp}(z) &=\frac{1}{4\TR\nf\seqav}\left[ \Pqgnlo-\Pqqlo\otimes\cftgnlo+\Pgglo\otimes\cftgnlo-b_0\cftgnlo\right]+2(\Pqglo\otimes\Gnlo)_{\wflpp} \,, \\
   \label{eq: PF2cwF3NLO}
P^{(1)}_{\ftcw \fk}(z) &=\frac{2(\eu-\ed)}{(A_d+A_u)(\ed+\eu)}\left[\Psingletnlo+\Pgqlo\otimes\cftgnlo\right]+2(\Pqglo\otimes\Gnlo)_{\fk} \,, \\
    \label{eq: PF2cwF2WdeltaNLO}
P^{(1)}_{\ftcw \wftwdelta}(z) &=\frac{(A_u-A_d)(\ed-\eu)}{(A_d+A_u)(\ed+\eu)}\left[\Psingletnlo +\Pgqlo\otimes\cftgnlo\right]+2(\Pqglo\otimes\Gnlo)_{\wftwdelta} \,, \\
\label{eq: PF2cwF3wNLO}
P^{(1)}_{\ftcw \fkw}(z) &=\frac{2(\ed-\eu)}{\ed+\eu}\left[\Psingletnlo+\Pgqlo\otimes\cftgnlo\right]+2 (\Pqglo\otimes\Gnlo)_{\fkw} \,, \\
\label{eq: PF2cwF2cwNLO}
P^{(1)}_{\ftcw \wftcw}(z) &= \Pplusnlo-b_0\cftqnlo \,.  
\end{align}

Convolutions $\cflqlo\otimes\Pqglo\otimes\Gnlo$ and $\cflglo\otimes\Gnlo$ appearing in the DGLAP evolution of $\fl$ in Eq.~\eqref{eq: FL DGLAP} can be written with 
\begin{align}
    (\cflqlo\otimes\Pqglo\otimes\Gnlo)_{\wft}=&-\frac{\cf}{4\nf\seqav}\Bigg\{ \frac{2\seqav}{(\ed+\eu)}\left[ \cflpsnlo+2(1-z)\otimes\cflpsnlo\right]+ \cflnsnlo+2(1-z)\otimes\cflnsnlo\nonumber\\
 -\frac{\cf}{\TR}&\left[ \cflgnlo+2(1-z)\otimes\cflgnlo\right] -\left[\cflqlo+2(1-z)\otimes\cflqlo\right]\otimes\left[\cftqnlo -\frac{\cf}{\TR}\cftgnlo\right] \Bigg\}\,, \\
    (\cflqlo\otimes\Pqglo\otimes\Gnlo)_{\wftp}=&-\frac{\cf^2}{8\TR\nf\seqav}\left\{\cflgnlo+2(1-z)\otimes\cflgnlo-\left[\cflqlo+2(1-z)\otimes\cflqlo\right]\otimes\cftgnlo \right\}
    \,, \\
    (\cflqlo\otimes\Pqglo\otimes\Gnlo)_{\wfl}=&\frac{1}{\cf}(\cflqlo\otimes\Pqglo\otimes\Gnlo)_{\wftp}\,, \\
    (\cflqlo\otimes\Pqglo\otimes\Gnlo)_{\wflp}=&-\frac{1}{\cf}(\cflqlo\otimes\Pqglo\otimes\Gnlo)_{\wftp}\,, \\
    (\cflqlo\otimes\Pqglo\otimes\Gnlo)_{\wflpp}=&\frac{1}{2\cf}(\cflqlo\otimes\Pqglo\otimes\Gnlo)_{\wftp}\,,\\
      (\cflqlo\otimes\Pqglo\otimes\Gnlo)_{\fk}= & -\frac{\cf}{4\nf}\frac{\eu-\ed }{(A_d+A_u)(\ed+\eu)}\left[ \cflpsnlo+2(1-z)\otimes\cflpsnlo \right] \,, \\ 
      (\cflqlo\otimes\Pqglo\otimes\Gnlo)_{\wftwdelta} =& -\frac{1}{2}(A_u-A_d)(\cflqlo\otimes\Pqglo\otimes\Gnlo)_{\fk} \,, \\  (\cflqlo\otimes\Pqglo\otimes\Gnlo)_{\fkw}=&-(A_d+A_u)(\cflqlo\otimes\Pqglo\otimes\Gnlo)_{\fk} \,, 
\end{align}
and
\begin{align}
    (\cflglo\otimes\Gnlo)_{\wft}=&-\frac{1}{2\nf\seqav}\Bigg\{ \frac{2\seqav}{(\ed+\eu)} \cflpsnlo+ \cflnsnlo  -\frac{\cf}{\TR} \cflgnlo -\cflqlo\otimes\left[\cftqnlo -\frac{\cf}{\TR}\cftgnlo\right]  \Bigg\}\,, \\
    (\cflglo\otimes\Gnlo)_{\wftp}=&-\frac{1}{2\nf\seqav}\Bigg\{ \frac{\cf}{2\TR} \cflgnlo -\frac{\cf}{2\TR}\cflqlo\otimes\cftgnlo  \Bigg\}\,, \\
    (\cflglo\otimes\Gnlo)_{\wfl}=&\frac{1}{\cf}(\cflglo\otimes\Gnlo)_{\wftp}\,, \\
    (\cflglo\otimes\Gnlo)_{\wflp}=&-\frac{1}{\cf}(\cflglo\otimes\Gnlo)_{\wftp}\,, \\
    (\cflglo\otimes\Gnlo)_{\wflpp}=&\frac{1}{2\cf}(\cflglo\otimes\Gnlo)_{\wftp} \,,\\
      (\cflglo\otimes\Gnlo)_{\fk}=& \frac{-1}{2\nf}\frac{\eu-\ed}{(A_d+A_u)(\ed+\eu)}\cflpsnlo \,,\\(\Pqglo\otimes\Gnlo)_{\wftwdelta}=& \frac{-1}{4\nf}\frac{(A_u-A_d)(\ed-\eu)}{(A_d+A_u)(\ed+\eu)}\cflpsnlo \,, \\  (\cflglo\otimes\Gnlo)_{\fkw}=&\frac{-1}{2\nf}\frac{\ed-\eu}{\ed+\eu}\cflpsnlo\,.
\end{align}
With these and by defining
\begin{equation}
    \label{eq:dFL pure singlet}
    P^{(1)}_{\fl \PSlo}(z) = 2\nf\seqav\cflqlo\otimes\Psingletnlo +2\nf\seqav\cflglo\otimes\Pgqnlo
+\seqav\Pqqlo\otimes\cflpsnlo+2\nf\seqav\Pgqlo\otimes\cflgnlo-b_0\seqav\cflpsnlo
\end{equation}
we can list the NLO evolution kernels for $\fl$:
\begin{align}
    \label{eq: PFLF2NLO}
P^{(1)}_{\fl \wft}(z) &= \frac{2}{\ed+\eu}P^{(1)}_{\fl \PSlo}(z)+\cflqlo\otimes\left[ \Pplusnlo -\frac{\cf}{\Tr}\Pqgnlo-\Pqqlo\otimes\left(\cftqnlo-\frac{\cf}{\TR}\cftgnlo \right)\right]\nonumber \\
&+2\nf\seqav (\cflqlo\otimes\Pqglo\otimes\Gnlo)_{\wft}  +2\nf\seqav\cflglo\otimes\left[ -\frac{\cf}{2\TR\nf\seqav}\Pggnlo-\frac{2}{\ed+\eu}\Pgqlo\otimes\left(\cftqnlo-\frac{\cf}{\Tr}\cftgnlo\right)\right]\nonumber \\&
+2\nf\seqav\Pgglo\otimes(\cflglo\otimes\Gnlo)_{\wft}-\frac{\cf}{\TR}\cflpsnlo\otimes\Pqglo \nonumber \\&
+\cflnsnlo\otimes\left[\Pqqlo -\frac{\cf}{\TR}\Pqglo\right] -\frac{\cf}{\TR}\cflgnlo\otimes\Pgglo 
-b_0\left[\cflnsnlo-\frac{\cf}{\TR}\cflgnlo \right]
\,, \\
    \label{eq: PFLF2pNLO}
P^{(1)}_{\fl \wftp}(z) &= \frac{\cf}{2\TR}\cflqlo\otimes\left[ \Pqgnlo-\Pqqlo\otimes\cftgnlo\right] +2\nf\seqav (\cflqlo\otimes\Pqglo\otimes\Gnlo)_{\wftp}  \nonumber \\&+\frac{\nf\seqav\cf}{\TR}\cflglo\otimes\left[\frac{1}{2\nf\seqav}\Pggnlo-\frac{2}{\ed+\eu}\Pgqlo\otimes\cftgnlo\right]
+2\nf\seqav\Pgglo\otimes(\cflglo\otimes\Gnlo)_{\wftp}\nonumber \\&+\frac{\cf}{2\TR}\Pqglo\otimes\left[\cflpsnlo+\cflnsnlo\right] 
+\frac{\cf}{2\TR}\Pgglo\otimes\cflgnlo 
-\frac{\cf b_0}{2\TR}\cflgnlo \,, \\
    \label{eq: PFLFLNLO}
P^{(1)}_{\fl \wfl}(z) &= \frac{1}{2\TR}\cflqlo\otimes\left[ \Pqgnlo-\Pqqlo\otimes\cftgnlo\right]+2\nf\seqav(\cflqlo\otimes\Pqglo\otimes\Gnlo)_{\wfl} \nonumber \\&
 +\frac{\nf\seqav}{\Tr}\cflglo\otimes\left[\frac{1}{2\nf\seqav}\Pggnlo-\frac{2}{\ed+\eu}\Pgqlo\otimes\cftgnlo \right] +2\nf\seqav\Pgglo\otimes(\cflglo\otimes\Gnlo)_{\wfl} \nonumber \\ &+\frac{1}{2\TR}\Pqglo\otimes\left[\cflpsnlo+\cflnsnlo \right] +\frac{1}{2\TR}\Pgglo\otimes\cflgnlo-\frac{b_0}{2\TR}\cflgnlo \,, 
\end{align}
\begin{align}
 \label{eq: PFLFLpNLO}
P^{(1)}_{\fl \wflp}(z) &= -\frac{1}{2\TR}\cflqlo\otimes\left[ \Pqgnlo-\Pqqlo\otimes\cftgnlo\right]+2\nf\seqav(\cflqlo\otimes\Pqglo\otimes\Gnlo)_{\wflp} \nonumber \\&
 -\frac{\nf\seqav}{\Tr}\cflglo\otimes\left[\frac{1}{2\nf\seqav}\Pggnlo-\frac{2}{\ed+\eu}\Pgqlo\otimes\cftgnlo \right] +2\nf\seqav\Pgglo\otimes(\cflglo\otimes\Gnlo)_{\wflp} \nonumber \\ &-\frac{1}{2\TR}\Pqglo\otimes\left[\cflpsnlo+\cflnsnlo \right] -\frac{1}{2\TR}\Pgglo\otimes\cflgnlo+\frac{b_0}{2\TR}\cflgnlo \,,\\
 \label{eq: PFLFLppNLO}
P^{(1)}_{\fl \wflpp}(z) &= \frac{1}{4\TR}\cflqlo\otimes\left[ \Pqgnlo-\Pqqlo\otimes\cftgnlo \right]+2\nf\seqav(\cflqlo\otimes\Pqglo\otimes\Gnlo)_{\wflpp} \nonumber \\&
 +\frac{\nf\seqav}{2\Tr}\cflglo\otimes\left[\frac{1}{2\nf\seqav}\Pggnlo-\frac{2}{\ed+\eu}\Pgqlo\otimes\cftgnlo \right] +2\nf\seqav\Pgglo\otimes(\cflglo\otimes\Gnlo)_{\wflpp} \nonumber \\ &+\frac{1}{4\TR}\Pqglo\otimes\left[\cflpsnlo+\cflnsnlo \right] +\frac{1}{4\TR}\Pgglo\otimes\cflgnlo-\frac{b_0}{4\TR}\cflgnlo \,, \\
   \label{eq: PFLF3NLO}
P^{(1)}_{\fl \fk}(z) &= \frac{\eu-\ed}{(A_d+A_u)(\ed+\eu)}P^{(1)}_{\fl \PSlo}(z)+2\nf\seqav(\cflqlo\otimes\Pqglo\otimes\Gnlo)_{\fk}\nonumber \\ &+2\nf\seqav\Pgglo\otimes (\cflglo\otimes\Gnlo)_{\fk} -\frac{(\eu-\ed)2\nf\seqav}{(A_d+A_u)(\ed+\eu)}\cflglo\otimes\Pgqlo\otimes\cfkqnlo  \,, \\
    \label{eq: PFLF2WdeltaNLO}
P^{(1)}_{\fl \wftwdelta}(z) &= \frac{(A_u-A_d)(\ed-\eu)}{2(A_d+A_u)(\ed+\eu)}P^{(1)}_{\fl \PSlo}(z)+2\nf\seqav(\cflqlo\otimes\Pqglo\otimes\Gnlo)_{\wftwdelta}\nonumber \\ &+2\nf\seqav\Pgglo\otimes (\cflglo\otimes\Gnlo)_{\wftwdelta} -\frac{(A_u-A_d)(\ed-\eu)}{2(A_d+A_u)(\ed+\eu)}2\nf\seqav\cflglo\otimes\Pgqlo\otimes\cftqnlo \,, \\
    \label{eq: PFLF3wNLO}
P^{(1)}_{\fl \fkw}(z) &= \frac{\ed-\eu}{\ed+\eu}P^{(1)}_{\fl \PSlo}(z)+2\nf\seqav(\cflqlo\otimes\Pqglo\otimes\Gnlo)_{\fkw} \nonumber \\ &+2\nf\seqav\Pgglo\otimes (\cflglo\otimes\Gnlo)_{\fkw}-\frac{\ed-\eu}{\ed+\eu}\cflglo\otimes\Pgqlo\otimes\cfkqnlo \,.
\end{align}

\bibliographystyle{JHEP-2modlong.bst}
\bibliography{refs}

\end{document}